\documentclass[10pt, a4paper, fleqn]{article}

\usepackage{amsmath}
\usepackage{natbib}
\usepackage{graphicx}
\usepackage{bm} 
\usepackage{amssymb} 
\usepackage{amsfonts} 
\usepackage{bbm} 
\usepackage{appendix}
\usepackage[explicit]{titlesec}
\usepackage[a4paper]{geometry} 
\usepackage{setspace}
\usepackage{microtype} 
\usepackage[colorinlistoftodos,disable]{todonotes} 
\usepackage{hyperref}
\usepackage{amsthm} 
\usepackage[all]{hypcap} 
\usepackage{multirow} 

\usepackage[nolists,nomarkers]{endfloat}

\usepackage[mathlines]{lineno}

\newcommand*\patchAmsMathEnvironmentForLineno[1]{%
  \expandafter\let\csname old#1\expandafter\endcsname\csname #1\endcsname
  \expandafter\let\csname oldend#1\expandafter\endcsname\csname end#1\endcsname
  \renewenvironment{#1}%
     {\linenomath\csname old#1\endcsname}%
     {\csname oldend#1\endcsname\endlinenomath}}%
\newcommand*\patchBothAmsMathEnvironmentsForLineno[1]{%
  \patchAmsMathEnvironmentForLineno{#1}%
  \patchAmsMathEnvironmentForLineno{#1*}}%
\AtBeginDocument{%
\patchBothAmsMathEnvironmentsForLineno{equation}%
\patchBothAmsMathEnvironmentsForLineno{align}%
\patchBothAmsMathEnvironmentsForLineno{flalign}%
\patchBothAmsMathEnvironmentsForLineno{alignat}%
\patchBothAmsMathEnvironmentsForLineno{gather}%
\patchBothAmsMathEnvironmentsForLineno{multline}%
}

\usepackage{sectsty}

\allsectionsfont{\centering}

\newtheorem{result}{Result}

\newcommand{\Ss}{\mathcal{S}} 
\newcommand{\Gg}{\mathcal{G}} 
\newcommand{\Cc}{\mathcal{C}} 
\newcommand{\Bb}{\mathcal{B}} 

\newcommand{\focal}{\bullet}
\newcommand{\deme}{\circ}

\newcommand{\dd}{\mathrm{d}} 

\DeclareMathOperator{\Esp}{E}

\newcommand{\dk}{d_k} 
\newcommand{\fk}{f_k} 
\newcommand{\Dfk}{\Delta f_k} 

\newcommand{\rc}{\kappa}
\newcommand{\cc}{\gamma}
\newcommand{\bb}{\beta}

\newcommand{\No}{N_{\mathrm{o}}}
\newcommand{\J}{O}

\begin{document}

\title{Relatedness and synergies of kind and scale in the evolution of helping}
\author{Jorge Pe\~na$^{1,*}$ \and Georg N\"oldeke$^{2}$ \and Laurent Lehmann$^{3}$}

\date{\today}

\maketitle

\onehalfspace

\vfill

\begin{itemize}
\item[$^{1}$] Department of Evolutionary Theory \\ Max Planck Institute for Evolutionary Biology \\ August-Thienemann-Str.~2, 24306 Pl\"on, Germany \\
e-mail: \href{mailto:pena@evolbio.mpg.de}{pena@evolbio.mpg.de}\\
\item[$^{2}$] Faculty of Business and Economics \\ University of Basel \\ Peter Merian-Weg 6, CH-4002 Basel, Switzerland \\
e-mail: \href{mailto:georg.noeldeke@unibas.ch}{georg.noeldeke@unibas.ch}\\
\item[$^{3}$] Department of Ecology and Evolution \\ University of Lausanne \\ Le Biophore,  CH-1015 Lausanne, Switzerland \\
e-mail: \href{mailto:laurent.lehmann@unil.ch}{laurent.lehmann@unil.ch}\\
\item[*] Corresponding author.
\end{itemize}

\newpage

\begin{abstract}

Relatedness and synergy affect the selection pressure on cooperation and altruism.
Although early work investigated the effect of these factors independently of each other, recent efforts have been aimed at exploring their interplay.
Here, we contribute to this ongoing synthesis in two distinct but complementary ways.
First, we integrate models of $n$-player matrix games into the direct fitness approach of inclusive fitness theory, hence providing a framework to consider synergistic social interactions between relatives in family and spatially structured populations.
Second, we illustrate the usefulness of this framework by delineating three distinct types of helping traits (``whole-group'', ``nonexpresser-only'' and ``expresser-only''), which are characterized by different synergies of kind (arising from differential fitness effects on individuals expressing or not expressing helping) and can be subjected to different synergies of scale (arising from economies or diseconomies of scale).
We find that relatedness and synergies of kind and scale can interact to generate nontrivial evolutionary dynamics, such as cases of bistable coexistence featuring both a stable equilibrium with a positive level of helping and an unstable helping threshold.
This broadens the qualitative effects of relatedness (or spatial structure) on the evolution of helping.

\smallskip
\noindent \textbf{Keywords.} evolution of helping, relatedness, synergy, inclusive fitness, evolutionary games

\end{abstract}

\newpage



\section{Introduction}
\label{sec:introduction}

Explaining the evolution of helping (cooperation and altruism) has been a main focus of research in evolutionary biology over the last fifty years (e.g., \citealp{Sachs2004,West2007}).
In this context, \citeauthor{Hamilton1964}'s seminal papers established the importance of relatedness (genetic assortment between individuals)
by showing that an allele for helping can be favored by natural selection as long as $-c+rb > 0$ is satisfied, where $c$ is the fitness cost to an average carrier from expressing the allele, $b$ is the fitness benefit to such a carrier stemming from a social partner expressing the allele, and $r$ is the relatedness between social partners \citep{Hamilton1964,Hamilton1964a,Hamilton1970}.
Additional factors, including different forms of reciprocity (i.e., conditional behaviors and responsiveness under multimove interactions, e.g., \citealp{Trivers1971,Axelrod1981})
and synergy (i.e., nonadditive effects of social behaviors on material payoffs, either positive or negative, e.g., \citealp{Queller1985,Sumpter2010}), modify the fitness costs and benefits in Hamilton's rule \citep{Axelrod1981,Day1997,Lehmann2006,Gardner2011,VanCleve2014}
and hence fundamentally influence the evolutionary dynamics of helping.

Because of their ubiquity, relatedness and synergy occupy a central role among the factors affecting the selection pressure on helping.
Both are clearly present in the cooperative enterprises of most organisms.
First, real populations are characterized by limited gene flow at least until the stage of offspring dispersal \citep{Clobert2001}, with the consequence that most social interactions necessarily occur between relatives of varying degree.
Second, social exchanges often feature at least one of two different forms of synergy, which we call in this article ``synergies of kind'' and ``synergies of scale''.

Synergies of kind (implicit in what \citealp{Queller2011} calls ``kind selection'') arise when the expression of a social trait benefits recipients in different ways, depending on whether or not (or more generally, to which extent) recipients express the social trait themselves.
A classical example of a positive synergy of kind is collective hunting \citep{Packer1988}, where the benefits of a successful hunt go to cooperators (hunters) but not to defectors (solitary individuals).
Examples of negative synergies of kind are eusociality in Hymenoptera, by which sterile workers help queens to reproduce \citep{Bourke1995}, and self-destructive cooperation in bacteria, where expressers lyse while releasing virulence factors that benefit nonexpressers \citep{Froehlich2000,Ackermann2008}.

Synergies of scale \citep{Corning2002} result from economies or diseconomies of scale in the production of a social good, so that the net effect of several individuals behaving socially can be more or less than the sum of individual effects.
For instance, enzyme production in microbial cooperation is likely to be nonlinear, as in the cases of invertase hydrolyzing disaccharides into glucose in the budding yeast \emph{Saccharomyces cerevisiae} \citep{Gore2009} or virulence factors triggering gut inflammation (and hence removal of competitors) in the pathogen \emph{Salmonella typhimurium} \citep{Ackermann2008}.
In the former case, the relationship between growth rate and glucose concentration in yeast has been reported to be sublinear, i.e., invertase production has diminishing returns or negative synergies of scale \citep[fig.~3.\emph{c}]{Gore2009}; in the latter case, the relationship between the level of expression of virulence factors and inflammation intensity appears to be superlinear, i.e., it exhibits increasing returns or positive synergies of scale \citep[fig.~2.\emph{d}]{Ackermann2008}.

Previous theoretical work has investigated the effects of relatedness and synergy on the evolution of helping either independently of each other or by means of simplified models that neglect crucial interactions between the two factors.
For instance, the effects of demography on relatedness and the scale of competition in family and spatially structured populations have often been explored under the assumption of additive payoff effects (e.g., \citealp{Taylor1992,Taylor2000,Lehmann2006b,Gardner2006}), while synergistic interactions have usually been investigated under the assumption that individuals are unrelated (e.g., \citealp{Motro1991,Leimar1998,Hauert2006a}).
In the cases where relatedness and synergy have been considered to operate in conjunction, it has been customary to model social interactions by means of a two-player Prisoner's Dilemma, modified by adding a synergy parameter $D$ to the payoff of mutual cooperation \citep{Grafen1979,Queller1984,Queller1985,Queller1992,Fletcher2006,Lehmann2006,Lehmann2006a,Ohtsuki2010,Gardner2011,Ohtsuki2012,Taylor2012,VanCleve2014}.
In this framework, $D>0$ (positive synergy) implies positive frequency-dependent selection, while $D<0$ (negative synergy) implies negative frequency-dependent selection.
The value of relatedness only matters in determining whether or not positive synergy leads to bistability, resp., whether or not negative synergy leads to coexistence.

Although illuminating in some aspects, such two-player models cannot capture patterns of synergy and resulting frequency dependence where positive (resp. negative) synergies of kind and negative (resp. positive) synergies of scale do not combine additively.
Such situations are however likely to be common in nature.
For example, collective hunting often features both positive synergies of kind and negative synergies of scale \citep{Packer1988}, while the production of virulence factors in \emph{S. typhimurium} features both negative synergies of kind and positive synergies of scale \citep{Ackermann2008}.
Models of two-player matrix games between relatives miss these patterns of synergy (and possible interactions between relatedness and synergy) because such games are linear, and only nonlinear games (which necessarily involve at least three-party interactions) can accommodate both negative and positive synergies without conflating them into a single parameter.
Although previous work has explored instances of $n$-player games between relatives (e.g., \citealp{Boyd1988,Eshel1988,Archetti2009,VanCleve2013,Marshall2014}) this has been done only for specific population or payoff structures, and hence not in a comprehensive manner.

In this article, we study the interplay between relatedness and synergies of kind and scale in models of $n$-player social interactions between relatives.
In order to do so, we first present a general framework that integrates $n$-player matrix games (e.g., \citealp{Kurokawa2009,Gokhale2010}) into the ``direct fitness'' approach \citep{Taylor1996,Rousset2004} of social evolution theory.
This framework allows us to deliver a tractable expression for the selection gradient (or gain function) determining the evolutionary dynamics, which differs from the corresponding expression for $n$-player games between unrelated individuals only in that ``inclusive gains from switching" rather than solely ``direct gains from switching" must be taken into account.

We then use the theoretical framework to investigate the interaction between relatedness, synergies of kind, and synergies of scale in the evolution of helping.
We show the importance of distinguishing between three different kinds of helping traits (which we call ``whole-group'', ``nonexpresser-only'' and ``expresser-only''), that are characterized by different types of synergies of kind (none for ``whole-group'', negative for ``nonexpresser-only'', positive for ``expresser-only''), and can be subjected to different synergies of scale.
Our analysis demonstrates that the interplay between relatedness and synergy can lead to patterns of frequency dependence, evolutionary dynamics, and bifurcations that cannot arise when considering synergistic interactions between unrelated individuals.
Thereby, our approach illustrates how relatedness and synergy combine nontrivially to affect the evolution of social behaviors.

\section{Modeling framework}
\label{sec:model}

\subsection{Population structure (demography)}

We consider a homogeneous haploid population subdivided into a finite and constant number of groups, each with a constant number $N \geq 1$ of adult individuals (see table \ref{table:notation} for a list of symbols).
The following events occur cyclically and span a demographic time period.
Each adult individual gives birth to offspring and then survives with a constant probability, so that individuals can be semelparous (die after reproduction) or iteroparous (survive for a number of demographic time periods).
After reproduction, offspring dispersal occurs.
Then, offspring in each group compete for breeding spots vacated by the death of adults so that exactly $N$ individuals reach adulthood in each group. 

Dispersal between groups may follow a variety of schemes, including the island model of dispersal \citep{Wright1931,Taylor1992}, isolation by distance \citep{Malecot1975,Rousset2004}, hierarchical migration \citep{Sawyer1983,Lehmann2012}, a model where groups split into daughter groups and compete against each other \citep{Gardner2006,Lehmann2006b,Traulsen2006}, and several variants of the haystack model (e.g., \citealp{Matessi1976,GodfreySmith2009}).
We leave the exact details of the life history unspecified, but assume that they fall within the scope of models of spatially homogeneous populations with constant population size (see \citealp[ch. 6]{Rousset2004}).

\subsection{Social interactions (games and payoffs)}
\label{subsec:social}

Each demographic time period, individuals interact socially by participating in a game between $n$ players.
Interactions can occur among all adults in a group ($n=N$), among a subset of such individuals ($n<N$) or among offspring before dispersal ($n > N$).
Individuals may either express a social behavior (e.g., cooperate in a Prisoner's Dilemma) or not (e.g., defect in a Prisoner's Dilemma). We denote these two possible actions by A (``cooperation") and B (``defection") and also refer to A-players as ``expressers" and to B-players as ``nonexpressers''.
The game is symmetric so that, from the point of view of a focal individual, any two co-players playing the same action are exchangeable.
We denote by $a_k$ the material payoff to an A-player when $k=0,1, \ldots,n-1$ co-players choose A (and hence $n-1-k$ co-players choose B).
Likewise, we denote by $b_k$ the material payoff to a B-player when $k$ co-players choose A.

We assume that individuals implement mixed strategies, i.e., they play A with probability $z$ (and hence play B with probability $1-z$).
The set of available strategies is then the interval $z \in[0,1]$.
At any given time only two strategies are present in the population: residents who play A with probability $z$ and mutants who play A with probability $z + \delta$.
Let us denote by $z_{\focal}$ the strategy (either $z$ or $z + \delta$) of a focal individual, and by $z_{\ell(\bullet)}$ the strategy of the $\ell$-th co-player of such focal.
The expected payoff $\pi$ to the focal is then
\begin{equation}
\label{eq:pi}
 \pi \left(z_{\focal},z_{1(\bullet)},z_{2(\bullet)},...,z_{n-1(\bullet)}\right) = \sum_{k=0}^{n-1} \phi_k\!\left(z_{1(\bullet)},z_{2(\bullet)},\ldots,z_{n-1(\bullet)}\right) \left[ z_\focal a_k + (1-z_\focal) b_k \right],
\end{equation}
where $\phi_k$ is the probability that exactly $k$ co-players play action A.
A first-order Taylor-series expansion about the average strategy $z_{\deme} = \sum^{n-1}_{\ell=1}z_{\ell(\bullet)}/(n-1)$ of co-players shows that, to first order in $\delta$,  the probability $\phi_k$ is given by a binomial distribution with parameters $n-1$ and $z_\deme$, i.e.,
\begin{equation}
\label{eq:phikapprox}
 \phi_k\!\left(z_{1(\bullet)},z_{2(\bullet)},\ldots,z_{n-1(\bullet)}\right) = \binom{n-1}{k} z_\deme^k (1-z_\deme)^{n-1-k} + O(\delta^2).
\end{equation}
Substituting~\eqref{eq:phikapprox} into~\eqref{eq:pi} and discarding second and higher order terms, we obtain
\begin{equation}
 \label{eq:piapprox}
 \pi\left(z_\focal,z_\deme\right) = \sum_{k=0}^{n-1} \binom{n-1}{k} z_\deme^k (1-z_\deme)^{n-1-k} \left[ z_\focal a_k + (1-z_\focal) b_k \right]
\end{equation}
for the payoff of a focal individual as a function of the focal's strategy $z_\focal$ and the average strategy $z_\deme$ of the focal's co-players (see also \citealp[p.~95]{Rousset2004} and \citealp[p.~85]{VanCleve2013}).

\subsection{Gain function and convergence stability}

Consider a population of residents playing $z$ in which a single mutant $z+\delta$ appears due to mutation, and denote by $\rho$ the fixation probability of the mutant.
We take the phenotypic selection gradient $\Ss = {\left(\dd \rho/\dd \delta \right)_{\delta=0}}$ as a measure of evolutionary success of the mutant (\citealp[p. 819]{Rousset2000};~\citealp[p. 17]{VanCleve2014a}); indeed, $\Ss > 0$ entails that the mutant has a fixation probability greater than neutral under weak selection ($|\delta| \ll 1$).
In order to evaluate the fixation probability, we assume that each demographic time period the material payoff to an individual determines its own fecundity (number of offspring produced before competition) or that of its parent (if interactions occur among offspring) by letting the average fecundity of an adult relative to a baseline be equal to the average payoff of a focal actor (i.e., payoffs from the game have ``fecundity effects'' as opposed to ``survival effects'', e.g., \citealp{Taylor2000}).
With this and our demographic assumptions, $\Ss$ is proportional to the ``gain function'' given by
\begin{equation}
\label{eq:Gz}
\Gg(z) = \underbrace{\frac{\partial \pi(z_\focal,z_\deme)}{\partial z_\focal} \Bigg|_{z_\focal=z_\deme=z}}_{\text{``direct'' effect, } -\Cc(z)} + \rc \underbrace{\frac{\partial \pi(z_\focal,z_\deme)}{\partial z_\deme} \Bigg|_{z_\focal=z_\deme=z}}_{\text{``indirect'' effect, }\Bb(z)} = - \Cc(z) + \rc \Bb(z)
\end{equation}
(see, e.g., \citealp[eq.~7]{VanCleve2013}).

Equation~\eqref{eq:Gz} shows that the gain function $\Gg(z)$ is determined by three components.
First, the ``direct'' effect  $-\Cc(z)$, that describes the change in average payoff of the focal resulting from the focal infinitesimally changing its own strategy.
Second, the ``indirect'' effect $\Bb(z)$, that describes the change in average payoff of the focal resulting from the focal's co-players changing their strategy infinitesimally.
Third, the indirect effect is weighted by the ``scaled relatedness coefficient'' $\rc$, which is a measure of relatedness between the focal individual and its neighbors, demographically scaled so as to capture the effects of local competition on selection \citep{Queller1994,Lehmann2010,Akcay2012}.
We discuss these three components of the gain function in more detail in the following section.

Knowledge of equation~\eqref{eq:Gz} is sufficient to characterize convergent stable strategies \citep{Eshel1981,Eshel1983,Taylor1989,Christiansen1991,Geritz1998,Rousset2004}.
In our context, candidate convergent stable strategies are either ``singular points'' (i.e., values $z^* \in (0,1)$ such that $\Gg(z^*)=0$), or the two pure strategies $z=1$ (always play A) and $z=0$ (always play B).
In particular, a singular point $z^*$ is convergent stable if $\dd \Gg(z)/\dd z|_{z=z^*} < 0$.
Regarding the endpoints, $z=1$ (resp. $z=0$) is convergent stable if $\Gg(1) > 0$ (resp. $\Gg(0) < 0$).
In this article we focus on convergence stability, and thus do not consider the possibility of disruptive selection, which can be ruled out by assuming that the evolutionary dynamics proceeds strictly through a sequence of mutant invasions and fixation events (e.g., the ``substitution process''; \citealp{Gillespie1991}; the ``trait substitution process''; \citealp{Metz1996}; or the ``trait substitution sequence''; \citealp{Champagnat2006}).

\subsection{Inclusive gains from switching}
\label{subsec:inclusivegains}

From equation~\eqref{eq:Gz}, the condition for a mutant to be favored by selection can be written as $-\Cc(z)+\rc \Bb(z) > 0$.
This can be understood as a scaled form of the marginal version of Hamilton's rule \citep{Lehmann2010} with $\Cc(z)$ corresponding to the marginal direct costs and $\Bb(z)$ to the marginal indirect benefits of expressing an increased probability of playing action A.
These marginal costs and benefits are not measured in terms of actual fitness (number of adult offspring, which are the units of measurement of $b$ and $c$ in Hamilton's rule as given in the introduction, see e.g., \citealp[p. 113]{Rousset2004}), but in terms of fecundity via payoffs in a game. 
The scaled relatedness coefficient $\rc$ is also not equal to the regression definition of relatedness present in the standard Hamilton's rule, except for special cases where competition is completely global \citep{Queller1994}.

The coefficient $\rc$ is a function of demographic parameters such as migration rate, group size, and vital rates of individuals or groups, but is independent of the evolving trait $z$ \citep{VanCleve2013}.
For instance, in the island model with overlapping generations, $\rc = 2s(1-m)/(N[2-m(1-s)]+2(1-m)s)$, where $m$ is the migration rate and $s$ is the probability of surviving to the next generation (\citealp[eq. A10]{Taylor2000}; \citealp[app. A2]{Akcay2012}).
In broad terms, we have (i) $\rc > 0$ for population structures characterized by positive assortment and relatively global competition, (ii) $\rc=0$ for infinitely large panmictic populations or for viscous populations with local competition exactly compensating for increased assortment of strategies \citep{Taylor1992}, and (iii) $\rc < 0$ for population structures characterized by negative assortment and/or very strong local competition (e.g., density-dependent competition occurs before dispersal).
Scaled relatedness coefficients have been evaluated for many life cycle conditions (see table 1 of \citealp{Lehmann2010}, table 1 of \citealp{VanCleve2013}, and references therein; see also app. \ref{sec:kappaapp} for values of $\kappa$ under different variants of the haystack model). 

In contrast to $\rc$, which  depends only on population structure, the other two components of the gain function are solely determined by the payoff structure of the social interaction.
In the following, we show how $-\Cc(z)$ and $\Bb(z)$ can be expressed in terms of the payoffs $a_k$ and $b_k$ of the game.
Doing so delivers an expression for $\Gg(z)$ that can be analyzed with the same techniques applicable for games between unrelated individuals.
This expression provides the foundation for our subsequent analysis.

Imagine a focal individual playing B in a group where $k$ of its co-players play A.
Suppose that this focal individual unilaterally switches its action to A while its co-players hold fixed their actions, thus changing its payoff from $b_k$ to $a_k$.
As a consequence, the focal experiences a ``direct gain from switching'' given by
\begin{equation}
\label{eq:directgains}
 d_k= a_k - b_k , \ k=0,1,\ldots,n-1 .
\end{equation}
At the same time, each of the focal's co-players playing A experiences a change in payoff given by $\Delta a_{k-1} = a_k-a_{k-1}$ and each of the focal's co-players playing B experiences a change in payoff given by $\Delta b_k = b_{k+1}-b_k$.
Hence, taken as a block, the co-players of the focal experience a change in payoff given by
\begin{equation}
\label{eq:indirectgains}
 e_k = k \Delta a_{k-1} + (n-1-k) \Delta b_k , \ k=0,1,\ldots,n-1,
\end{equation}
where we let $a_{-1}= b_{n+1} = 0$ for mathematical convenience.
From the perspective of the focal, this change in payoffs represents an ``indirect gain from switching'' the focal obtains if co-players are related.

In appendix \ref{sec:gzapp}, we show that the partial derivatives appearing in~\eqref{eq:Gz} can be expressed as expected values of the direct and indirect gains from switching, so that the direct and indirect effects are respectively given by
\begin{equation}
 \label{eq:Cz}
 -\Cc(z) = \sum_{k=0}^{n-1} \binom{n-1}{k} z^k (1-z)^{n-1-k} d_k ,
\end{equation}
and
\begin{equation}
 \label{eq:Bz}
 \Bb(z) = \sum_{k=0}^{n-1} \binom{n-1}{k} z^k (1-z)^{n-1-k} e_k.
\end{equation}

Hence, defining the ``inclusive gains from switching'' as
\begin{equation}
\label{eq:inclusivegains}
 f_k= d_k + \rc e_k, \ k=0,1,\ldots,n-1,
\end{equation}
the gain function can be written as the expected value of the inclusive gains from switching:
\begin{equation}
\label{eq:Gzbinom}
 \Gg(z) = \sum_{k=0}^{n-1} \binom{n-1}{k} z^k (1-z)^{n-1-k} f_k .
\end{equation}

An immediate consequence of equation~\eqref{eq:Gzbinom} is that matrix games between relatives are mathematically equivalent to ``transformed'' games between unrelated individuals, where ``inclusive payoffs'' take the place of standard, or personal, payoffs.
Indeed, consider a game in which a focal playing A (resp. B) obtains payoffs
\begin{eqnarray}
\label{eq:payofft-1} a_k' & =  a_k + \rc \left[k a_k + (n-1-k) b_{k+1}\right], \ k=0,1,\ldots,n-1  \\
\label{eq:payofft-2} b_k' & = b_k + \rc \left[k a_{k-1} + (n-1-k) b_{k}\right], \ k=0,1,\ldots,n-1
\end{eqnarray}
when $k$ of its co-players play A.
Using equations~\eqref{eq:directgains}--\eqref{eq:indirectgains} we can rewrite equation~\eqref{eq:inclusivegains} as $f_k = a_k' - b_k'$, so that the inclusive gains from switching are identical to the direct gains from switching in a game with payoff structure given by equations \eqref{eq:payofft-1}--\eqref{eq:payofft-2}.
The payoffs $a_k'$ (resp. $b_k'$) can be understood as inclusive payoffs consisting of the payoff obtained by a focal playing A (resp. B) plus $\kappa$ times the sum of the payoffs obtained by its co-players.

This observation has two relevant consequences.
First, the results developed in \cite{Pena2014} for nonlinear $n$-player matrix games between unrelated individuals, which are based on the observation that the right side of~\eqref{eq:Gzbinom} is a polynomial in Bernstein form \citep{Farouki2012}, also apply here, provided that (i) the inclusive gains from switching $\fk$ are used instead of the standard (direct) gains from switching $\dk$ in the formula for the gain function, and (ii) the concept of evolutionary stability is read as meaning convergence stability.
For a large class of games, these results allow to identify convergence stable points from a direct inspection of the sign pattern of the inclusive gains from switching $\fk$. Second, we may interpret the effect of relatedness on selection as inducing the payoff transformation $a_k \rightarrow a_k'$, $b_k \rightarrow b_k'$.
For $n=2$, this payoff transformation is the one hinted at by \cite{Hamilton1971} and later often discussed in the theoretical literature \citep{Grafen1979,Hines1979,Day1998}, namely
\begin{equation*}
\label{eq:payofftransform}
 \left( \begin{array}{cc}
a_0' & a_1' \\
b_0' & b_1' \end{array} \right)
=
\left( \begin{array}{cc}
a_0 + \rc b_1 & (1+\rc)a_1 \\
(1+\rc)b_0 & b_1 + \rc a_0 \end{array} \right)
\, ,
\end{equation*}
where the payoff of the focal is augmented by adding $\kappa$ times the payoff of the co-player.

\section{Evolutionary dynamics of three kinds of helping traits}
\label{sec:results}

Throughout the following we assume that each A-player incurs a payoff cost $\cc > 0$ in order for a social good to be produced (e.g., harvested food, nest defense, or help directed to others).
The benefits of the social good are accrued by a subset of individuals in the group that we call ``recipients''.
Each recipient obtains a benefit $\bb_j > 0$ when there are $j>0$ expressers in the group, and no benefit is produced if no individual expresses the social trait ($\bb_0 = 0$).
The benefit is increasing in the number of expressers, that is, the ``incremental benefit" $\Delta \bb_j = \bb_{j+1} - \bb_{j}$ is positive ($\Delta \bb_j > 0$).

Synergies of scale are characterized by the properties of the incremental benefits.
In the absence of synergies of scale, each additional expresser increases the benefit by the same amount so that $\Delta \bb_j$ is constant, implying that $\bb_j$ is linear in $j$. With negative synergies of scale, $\Delta \bb_j$ is decreasing in $j$, whereas positive synergies of scale arise when $\Delta \bb_j$ is increasing in $j$.
To illustrate the effects of synergies of scale on the evolutionary dynamics of the social trait, we will consider the special case in which incremental benefits are given by the geometric sequence $\Delta \bb_j = \bb \lambda^j$ for some $\bb > 0$ and $\lambda > 0$, so that benefits are given by
\begin{equation}
\label{eq:geombenefits}
\bb_j = \bb \sum_{\ell = 0}^{j-1} \lambda^\ell.
\end{equation}
With geometric benefits,  synergies of scale are absent when $\lambda = 1$, negative when $\lambda < 1$, and positive when $\lambda > 1$.

\begin{figure}[t]
\centering
\includegraphics[width=\textwidth]{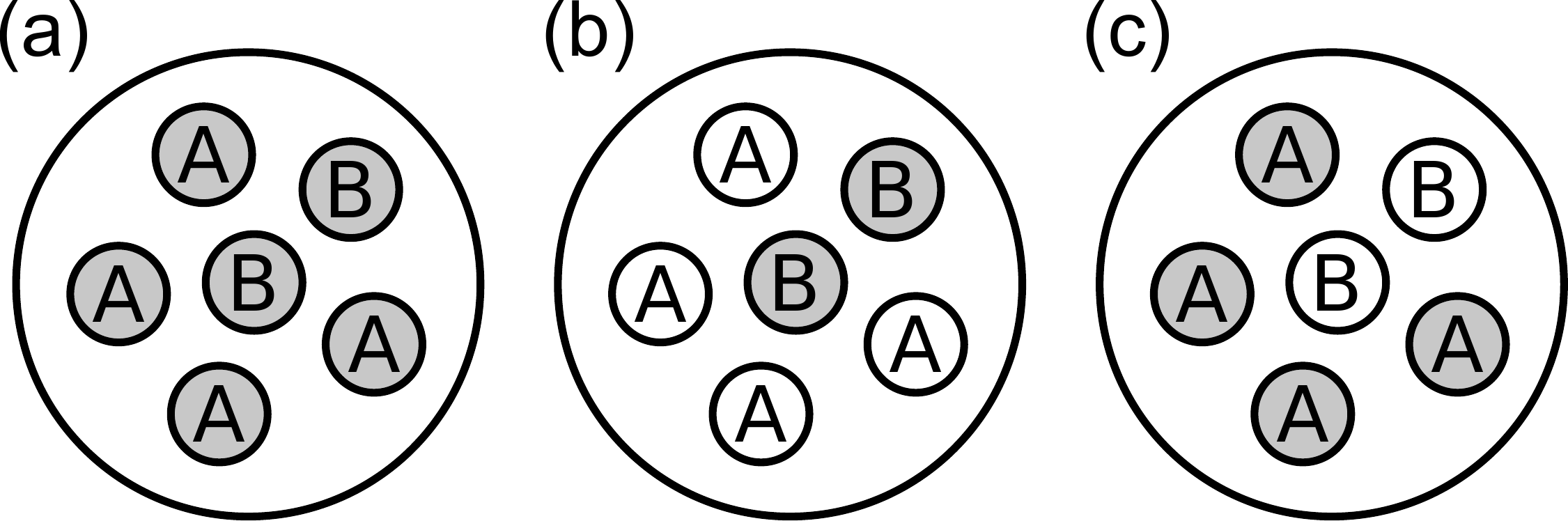}
\caption{Three kinds of social traits.
Expressers (As) provide a social good at a personal cost, nonexpressers (Bs) do not.
The set of recipients (filled circles) of the social good depends on the particular kind of social interaction.
\emph{a}, Whole-group traits.
\emph{b}, Nonexpresser-only traits.
\emph{c}, Expresser-only traits.
}
\label{fig:threekinds}
\end{figure}

We distinguish three kinds of social traits according to which individuals are recipients and thus benefit from the expression of the social behavior: (i) ``whole-group'' (benefits accrue to all individuals in the group, fig.~\ref{fig:threekinds}.\emph{a}), (ii) ``nonexpresser-only'' (benefits accrue only to nonexpressers, fig.~\ref{fig:threekinds}.\emph{b}),
and (iii) ``expresser-only'' (benefits accrue only to expressers, fig.~\ref{fig:threekinds}.\emph{c}).
For whole-group traits there are no synergies of kind: benefits accrue to all individuals irrespective of their kind, i.e., whether they are expressers or nonexpressers.
In contrast, nonexpresser-only traits feature negative synergies of kind, whereas expresser-only traits feature positive synergies of kind.
These differences are reflected in different payoff structures for the corresponding $n$-player games, resulting in different direct, indirect, and inclusive gains from switching (see table \ref{table:threekinds}).

A classical example of a whole-group trait is the voluntary provision of public goods \citep{Samuelson1954}.
In this case, the expressed social behavior consists in the production of a good available to others and hence exploitable by nonproducing cheats (nonexpressers).
Well-known instances of public-goods cooperation are sentinel behavior in animals \citep{MaynardSmith1965,Clutton-Brock1999}, and the secretion of extracellular products \citep{Velicer2003,West2007}, such as sucrose-digestive enzymes \citep{Greig2004,Gore2009}, in social bacteria.

The most prominent social behavior matching our definition of a nonexpresser-only trait is altruistic self-sacrifice, which happens when individuals expressing the social behavior sacrifice themselves (or their reproduction) to benefit nonexpressers \citep{Frank2006,West2006}.
Sterile castes in eusocial insects \citep{Bourke1995}, and bacteria lysing while releasing toxins \citep{Froehlich2000} or virulence factors \citep{Ackermann2008} that benefit other bacteria provide some examples of altruistic self-sacrifice in nature.

Expresser-only traits have been discussed under the rubrics of ``synergistic'' \citep{Queller1984,Queller1985,Leimar1998} and ``greenbeard'' \citep{Guilford1985,Gardner2010,Queller2011} effects, and conceptualized as involving ``rowing''(\citealp[p. 261-262]{MaynardSmith1995}) or ``stag hunt'' \citep{Skyrms2004} games.
Often cited examples include collective hunting \citep{Packer1988}, foundresses cooperating in colony establishment \citep{Bernasconi1999}, aposematic (warning) coloration \citep{Queller1984,Queller1985,Guilford1988}, and the \emph{Ti} plasmid in the bacterial pathogen \emph{Agrobacterium tumefaciens}, which induces its plant host to produce opines, a food source that can be exploited only by bacteria bearing the plasmid (\citealp[p. 218]{Dawkins1999}, \citealp{White2007}).
In each of these examples, the social good accrues only to partners expressing the trait, either because of a greater tendency to group and interact or because of the action of an emergent recognition system discriminating expressers from nonexpressers.

For all three kinds of social traits, the indirect gains from switching are always nonnegative ($e_k \geq 0$ for all $k$) and hence the indirect effect $\Bb(z)$ is nonnegative for all $z$.
This implies that we deal with helping traits at the level of payoffs and that increasing $\rc$ never leads to less selection for expressing the social behavior.
Due to their different ways of defining recipients, however, each social trait is characterized by a social dilemma with structurally different payoff, direct gain, and indirect gains from switching.
For nonexpresser-only traits, the direct gains from switching are always negative ($d_k < 0$ for all $k$) and thus expressing the social behavior is also payoff altruistic ($-\Cc(z)<0$ and $\Bb(z) \geq 0$ for all $z$).
For whole-group and expresser-only traits, expressing the social behavior is not necessarily altruistic, depending on how the cost $\cc$ compares to benefits ($\bb_{k+1}$, expresser-only traits) or incremental benefits ($\Delta \bb_k$, whole-group traits).

Before turning to the analysis, we note that a fourth class of social traits is sometimes also distinguished in the literature, namely ``other-only'' traits where the benefits accrue to all other individuals in the group, but not to the focal expresser itself \citep{Pepper2000}.
Other-only traits, as whole-group traits, lack synergies of kind, and hence the effects of relatedness on the evolutionary dynamics are qualitatively similar for whole-group and other-only traits.
We discuss this latter case in more detail in section \ref{subsec:connections} and relegate the formal analysis, which is similar to the one for whole-group traits, to appendix \ref{sec:otherapp}. 

\subsection[No synergies of scale]{No synergies of scale}
\label{subsec:linear}

To isolate the effects of synergies of kind, we begin our analysis with the case in which synergies of scale are absent, that is, benefits take the linear form $\bb_j = \beta j$ ($\lambda = 1$ in eq.~\eqref{eq:geombenefits}).
The resulting expressions for the inclusive gains from switching and the gain functions for the three different social traits are shown in table \ref{table:linear}.
In each case, the gain function can be written as
\begin{equation*} \label{eq:Gzlinear}
 \Gg(z) = (n-1)\left[- C + \rc B + (1+\rc) D z\right],
\end{equation*}
where the parameter $C > 0$ may be thought of as the ``effective cost'' per co-player of expressing the social trait when none of the co-players expresses the social trait.
We have $C=\gamma/(n-1)$ when a focal expresser is not among the recipients (nonexpresser-only traits) and  $C= (\gamma - \beta)/(n-1)$ otherwise (whole-group and expresser-only traits).
The parameter $B \ge 0$ measures the incremental benefit accruing to each co-player of a focal expresser when none of the co-players expresses the social trait.
We thus have $B=0$ for expresser-only traits and $B=\beta$ otherwise.
Finally, $D$ measures synergies of kind and is thus null for whole-group traits ($D=0$), negative for nonexpresser-only traits ($D = - \beta$) and positive for expresser-only traits ($D = \beta$).

In the absence of synergies of kind ($D=0$, whole-group traits) selection is frequency independent and defection dominates cooperation ($z=0$ is the only convergence stable strategy) if $- C + \rc B < 0$ holds, whereas cooperation dominates defection ($z=1$ is the only convergence stable strategy) if $- C + \rc B > 0$ holds.

With negative synergies of kind ($D < 0$), there is negative frequency-dependent selection.
Defection dominates cooperation if $-C + \rc B \leq 0$ holds, whereas cooperation dominates defection if $- C + \rc B + (1 + \rc) D \geq 0$ holds.
If $- C + \rc B + (1 + \rc) D < 0 < - C + \rc B$ holds, both $z=0$ and $z=1$ are unstable and the singular point
\begin{equation}
 \label{eq:ESSlinear}
 z^* = \frac{C-\rc B}{(1+\rc)D}
\end{equation}
is stable.

With positive synergies of kind ($D > 0$), there is positive frequency-dependent selection.
Defection dominates cooperation if $- C + \rc B + (1 + \rc) D \leq 0$ holds, whereas cooperation dominates defection if  $- C + \rc B \geq 0$ holds.
When $- C + \rc B < 0 < - C + \rc B + (1 + \rc) D$, there is bistability: both $z=0$ and $z=1$ are stable and $z^*$ is unstable.

This analysis reveals three important points.
First, in the absence of synergies of scale the gain function is linear in $z$, which allows for a straightforward analysis of the evolutionary dynamics for all three kinds of social traits.
Second, because of the linearity of the gain function, the evolutionary dynamics of such games fall into one of the four classical dynamical regimes arising from $2 \times 2$ games,
namely (i) A dominates B, (ii) B dominates A, (iii) coexistence, and (iv) bistability (see, e.g., \citealp[section 2.2]{Cressman2003}).
Third, which of these dynamical regimes arises is determined by the interaction of relatedness with synergies of kind in a straightforward fashion. For all traits, defection dominates cooperation when relatedness is low.
For whole-group traits, high values of relatedness imply that cooperation dominates defection.
For nonexpresser-only and expresser-only traits, high relatedness also promotes cooperation, leading to either the coexistence of expressers and nonexpressers (nonexpresser-only traits) or to bistability (expresser-only traits).

\subsection[Whole-group traits with synergies of scale]{Whole-group traits with synergies of scale}
\label{subsec:whole}

For whole-group traits there are no synergies of kind, but either positive or negative synergies of scale may arise.
How do such synergies of scale change the evolutionary dynamics of whole-group helping?
Substituting the inclusive gains from switching given in table \ref{table:threekinds} into equation~\eqref{eq:Gzbinom} shows that the gain function for whole-group traits is given by
\begin{equation}
\label{eq:Gzwhole}
 \Gg(z) = \sum_{k=0}^{n-1} \binom{n-1}{k} z^k (1-z)^{n-1-k} \left\{ - \cc + \left[1 + \rc (n-1)\right]\Delta \bb_k \right\}.
\end{equation}
Since the incremental benefit satisfies $\Delta \bb_k > 0$ for all $k$, the gain function~\eqref{eq:Gzwhole} is negative for $\kappa \leq -1/(n-1)$.
In this case, defection dominates cooperation and $z=0$ is the only stable point. Hence, we consider the case $\rc > -1/(n-1)$ throughout the following.

If synergies of scale are negative ($\Delta \bb_k$ decreasing in $k$), the direct gains ($d_k$) indirect gains ($e_k$) and inclusive gains ($f_k$) from switching are all decreasing in $k$.
This implies that $-\Cc(z)$, $\Bb(z)$ and $\Gg(z)$ are all decreasing in $z$ (cf.~\citealp[remark 3]{Pena2014}).
Similarly, if synergies of scale are positive ($\Delta \bb_k$ increasing in $k$), $d_k$, $e_k$ and $f_k$ are all increasing in $k$ and hence $-\Cc(z)$, $\Bb(z)$ and $\Gg(z)$ are all increasing in $z$.
In both cases the evolutionary dynamics are easily characterized by applying the results for public goods games with constant costs from \citet[section 4.3]{Pena2014}: with negative synergies of scale, defection dominates cooperation (so that $z = 0$ is the only convergent stable strategy) if $\cc \ge [1 +\rc (n-1)] \Delta \bb_0$, whereas cooperation dominates defection if $\cc \le [1 + \rc (n-1)] \Delta \bb_{n-1}$ holds.
If $[1 + \rc (n-1)] \Delta \bb_{n-1} < \cc < [1 +\rc (n-1)] \Delta \bb_0$ holds, there is coexistence: both $z=0$ and $z=1$ are unstable and there is a unique stable interior point $z^*$.
With positive synergies of scale, defection dominates cooperation if $\cc \ge [1 + \rc (n-1)] \Delta \bb_{n-1}$, whereas cooperation dominates defection if $\cc \le [1 +\rc (n-1)] \Delta \bb_0$.
If $[1 +\rc (n-1)] \Delta \bb_0 < \cc < [1 + \rc (n-1)] \Delta \bb_{n-1}$ holds, there is bistability: both $z=0$ and $z=1$ are stable and there is a unique, unstable interior point $z^*$ separating the basins of attraction of these two stable strategies.
These results resemble those for the cases in which there are no synergies of scale (section \ref{subsec:linear}), but negative, resp. positive synergies of kind are present.
In particular, it is again the case that the evolutionary dynamics fall into one of the four classical dynamical regimes arising from $2 \times 2$ games.

The effect of relatedness on the evolution of whole-group traits can be better grasped by noting that multiplying and dividing \eqref{eq:Gzwhole} by $1+\rc (n-1)$, we obtain
\begin{equation}
\label{eq:Gzwhole2}
 \Gg(z) = \left[ 1+\rc (n-1) \right] \sum_{k=0}^{n-1} {n-1 \choose k} z^k (1-z)^{n-1-k} \left( - \tilde \cc + \Delta \bb_k \right),
\end{equation}
where $\tilde \cc = \cc/[1+\rc (n-1)]$.
Equation~\eqref{eq:Gzwhole2} is (up to multiplication by a positive constant) equivalent to the gain function of a public goods game between unrelated individuals with payoff cost $\tilde \cc$ for producing the public good, which has been analyzed under different assumptions on the shape of the benefit sequence \citep{Motro1991,Bach2006,Hauert2006a,Pena2014}. Hence, relatedness can be conceptualized as affecting only the cost of cooperation,
while leaving synergies of scale and patterns of frequency dependence unchanged.

As a concrete example, consider the case of geometric benefits \eqref{eq:geombenefits} with $\lambda \neq 1$ (see table \ref{table:resultsfourkinds} for a summary of the results and app. \ref{sec:wholeapp} for a derivation).
We find that there are two critical cost-to-benefit ratios
\begin{equation}
 \label{eq:epsilonwhole}
 \varepsilon = \text{min}\left(1+ \rc (n-1), \lambda^{n-1} [1+ \rc (n-1)]\right) \text{ and }  \vartheta = \text{max}\left(1+ \rc (n-1), \lambda^{n-1} [1+ \rc (n-1)]\right) ,
\end{equation}
such that for small costs ($\cc/\beta \leq \varepsilon$) cooperation dominates defection ($z=1$ is the only stable point) and for large costs ($\cc/\beta \geq \vartheta$) defection dominates cooperation ($z=0$ is the only stable point).
For intermediate costs ($\varepsilon < \cc/\bb < \vartheta$), there is a singular point given by
\begin{equation}
\label{eq:zwholegroup}
z^*= \frac{1}{1-\lambda} \left[ 1-\left(\frac{\cc}{\bb\left[1+\rc (n-1)\right]}\right)^{\frac{1}{n-1}} \right] ,
\end{equation}
such that the evolutionary dynamics are characterized by coexistence if synergies of scale are negative ($\lambda<1$) and by bistability if synergies of scale are positive ($\lambda>1$).
It is clear from equation \eqref{eq:epsilonwhole} that, for a given cost-to-benefit ratio $\cc/\bb$, increasing relatedness makes larger (resp. smaller) the region in the parameter space where cooperation (resp. defection) dominates.
Moreover, and from equation~\eqref{eq:zwholegroup}, $z^*$ is an increasing (resp. decreasing) function of $\rc$ when $\lambda < 1$ (resp. $\lambda > 1$), meaning that the proportion of individuals cooperating at a stable interior point (resp. the size of the basin of attraction of the fully cooperative equilibrium) increases as a function of $\rc$ (see fig. \ref{fig:resultsfourkinds}.\emph{a} and \ref{fig:resultsfourkinds}.\emph{d}).

\subsection{Nonexpresser-only traits with synergies of scale}
\label{subsec:nonexpressersonly}

For nonexpresser-only traits, synergies of kind are negative.
In the absence of synergies of scale, and as discussed in section \ref{subsec:linear}, this implies negative frequency dependence.
To investigate how positive or negative synergies of scale change this baseline scenario, we focus on the case in which relatedness is nonnegative ($\rc \ge 0$).

From the formulas for $d_k$ and $e_k$ given in table \ref{table:threekinds}, it is clear that, independently of any synergies of scale, the direct gains from switching $d_k$ are decreasing in $k$.
Hence, the direct effect $-\Cc(z)$ is negative frequency-dependent.
When synergies of scale are negative, the indirect gains from switching $e_k$ are also decreasing in $k$, implying that the indirect effect $\Bb(z)$ is also negative frequency-dependent and that the same is true for the gain function $\Gg(z) = - \Cc(z) + \kappa \Bb(z)$. Hence, negative synergies of scale lead to evolutionary dynamics that are qualitatively identical to those arising when synergies of scale are absent: for low relatedness, defection dominates cooperation, and for sufficiently high relatedness, a unique interior stable equilibrium appears (see app. \ref{subsec:nonexpressersappconcave} and fig. \ref{fig:resultsfourkinds}.\emph{b}).

When synergies of scale are positive, the indirect gains from switching $e_k$ may still be decreasing in $k$ because the incremental gain $\Delta \bb_k$ accrues to a smaller number of recipients ($n-1-k$) as $k$ increases.
In such a scenario, always applicable when $n=2$, the evolutionary dynamics are again qualitatively identical to those arising when synergies of scale are absent.
A different picture can emerge if $n> 2$ holds and synergies of scale are not only positive, but also sufficiently strong.
Then, the indirect gains from switching may be unimodal (first increasing, then decreasing) in $k$, implying \citep{Pena2014} that the indirect benefit $\Bb(z)$ is similarly unimodal, featuring positive frequency dependence for small $z$ and negative frequency dependence for large $z$.
Depending on the value of relatedness, which modulates how the frequency dependence of $\Bb(z)$ interacts with that of  $\Cc(z)$, this can give rise to evolutionary dynamics different from those possible without synergies of scale, discussed in section \ref{subsec:linear}.

For a concrete example of such evolutionary dynamics, consider the case of geometric benefits \eqref{eq:geombenefits} with $\lambda > 1$ (see table \ref{table:resultsfourkinds} for a summary of results, app. \ref{subsec:nonexpressersappgeometric} for their derivation and fig. \ref{fig:resultsfourkinds}.\emph{e} for an illustration).
In this case, the evolutionary dynamics for $\rc \geq 0$ and $n > 2$ depend on the critical value
\begin{equation}
 \label{eq:varrho}
 \varrho = \frac{1+\rc (n-1)}{\rc (n-2)},
\end{equation}
and on the two critical cost-to-benefit ratios
\begin{equation}
 \label{eq:zeta}
 \zeta = \rc (n-1) ,\qquad\text{and}\qquad  \eta = \frac{1}{\lambda-1}\left[ 1 + \lambda \rc \left( \frac{(n-2)\lambda \rc}{1+\rc (n-1)} \right)^{n-2} \right] ,
\end{equation}
which satisfy $\varrho > 1$ and $\zeta < \eta$.

With these definitions our results can be stated as follows.
For $\lambda \leq \varrho$ the dynamical outcome depends on how the cost-to-benefit ratio $\cc/\bb$ compares to $\zeta$.
If $\cc/\bb \geq \zeta$ (high costs), defection dominates cooperation, while if $\cc/\bb < \zeta$ (low costs), there is coexistence.
For $\lambda > \varrho$, the dynamical outcome also depends on how the cost-to-benefit ratio $\cc/\bb$ compares to $\eta$.
If $\cc/\bb \geq \eta$ (high costs), defection dominates cooperation.
If $\cc/\bb \leq \zeta$ (low costs), we have coexistence, with the stable singular point $z^*$ satisfying $z^*>\hat{z}$
where
\begin{equation}
\label{eq:hatznonexpressers}
 \hat{z}=\frac{\rc \left[(n-2)\lambda - (n-1) \right]-1}{[1+\rc (n-1)](\lambda-1)} .
\end{equation}

In the remaining case ($\zeta < \cc/\bb < \eta$, intermediate costs) the dynamics are characterized by bistable coexistence, with $z = 0$ stable, $z=1$ unstable, and two singular points $z_\textrm{L}$ (unstable) and $z_\textrm{R}$ (stable) satisfying $0 < z_\textrm{L} < \hat{z} < z_\textrm{R} < 1$.
Numerical values for $z_\textrm{L}$ (resp. $z_\textrm{R}$) can be obtained by searching for roots of $\Gg(z)$ in the interval $(0,\hat{z})$ (resp. $(\hat{z},1)$), as we illustrate in figure \ref{fig:resultsfourkinds}.\emph{e}.

It is evident from the dependence of $\varrho$, $\zeta$, and $\eta$ on $\rc$ that relatedness plays an important role in determining the stable level(s) of expression of helping.
As $\rc$ increases, the regions of the parameter space where some non-zero level of expression of helping is stable expand at the expense of the region of dominant non-expression.
This is so because $\zeta$ and $\eta$ are increasing functions of $\rc$ and $\varrho$ is a decreasing function of $\rc$.
Moreover, inside these regions the stable non-zero probability of expressing helping increases with $\rc$ (see fig. \ref{fig:resultsfourkinds}.\emph{b} and \ref{fig:resultsfourkinds}.\emph{e}).
Three cases can be however distinguished as for the effects of increasing $\rc$ when starting from a point in the parameter space where $z=0$ is the only stable point.
First, $z=0$ can remain stable irrespective of the value of relatedness, which characterizes high cost-to-benefit ratios.
Second, the system can undergo a transcritical bifurcation as $\rc$ increases, destabilizing $z=0$ and leading to the appearance of a unique stable interior point (fig. \ref{fig:resultsfourkinds}.\emph{b}).
This happens when $\lambda$ and $\cc/\bb$ are relatively small.
Third, there is a range of intermediate cost-to-benefit ratios such that, for sufficiently large values of $\lambda$, the system undergoes a saddle-node bifurcation, whereby two singular points ($z_\textrm{L}$, unstable, and $z_\textrm{R}$, stable) appear (fig. \ref{fig:resultsfourkinds}.\emph{e}).
In this latter case, positive synergies of scale are strong enough to interact with negative synergies of kind and relatedness in a nontrivial way.

\subsection{Expresser-only traits with synergies of scale}
\label{subsec:expressersonly}

For expresser-only traits, and independently of any synergies of scale, the direct gains from switching $d_k$ (cf. table \ref{table:threekinds}) are increasing in $k$, implying that the direct effect $-\Cc(z)$ is positive frequency-dependent.
When synergies of scale are positive, the indirect gains from switching $e_k$ are also increasing in $k$, so that the indirect effect $\Bb(z)$ is also positive frequency-dependent.
Focusing on the case of nonnegative relatedness ($\rc \ge 0)$ this ensures that, just as when synergies of scale are absent, the gain function $\Gg(z)$ is positive frequency-dependent.
Hence, the evolutionary dynamics are qualitatively identical to those arising from linear benefits: for low relatedness, defection dominates cooperation, and for high relatedness, there is bistability, with the basins of attraction of the two pure equilibria $z=0$ and $z=1$ being separated by a unique interior unstable point (see app. \ref{subsec:expressersappconvex} and fig. \ref{fig:resultsfourkinds}.\emph{f}).

When synergies of scale are negative, the indirect gains from switching $e_k$ may still be increasing in $k$ because the incremental gain $\Delta \bb_k$ accrues to a larger number of recipients as $k$ increases. In such a scenario, always applicable when $n=2$, the evolutionary dynamics are again qualitatively identical to those arising when synergies of scale are absent.
A different picture can emerge if $n> 2$ holds and synergies of scale are not only negative, but also sufficiently strong.
In this case, $\Bb(z)$ can be negative frequency-dependent for some $z$, and hence (for sufficiently high values of $\kappa$) also $\Gg(z)$.
Similarly to the case of nonexpresser-only traits with positive synergies of scale, this can give rise to patterns of frequency dependence that go beyond the scope of helping without synergies of scale.

To illustrate this, consider the case of geometric benefits \eqref{eq:geombenefits} with $\lambda < 1$, $\rc \geq 0$, and $n>2$ (see table \ref{table:resultsfourkinds} for a summary of results, app. \ref{subsec:expressersappgeometric} for proofs and fig. \ref{fig:resultsfourkinds}.\emph{c} for an illustration).
Defining the critical value
\begin{equation}
 \label{eq:xi}
 \xi = \frac{\rc (n-2)}{1+\rc (n-1)},
\end{equation}
and the two critical cost-to-benefit ratios
\begin{equation}
 \label{eq:varsigma}
 \varsigma = \frac{1-\lambda^n}{1-\lambda} +\rc (n-1) \lambda^{n-1}, \qquad\text{and}\qquad  \tau=\frac{1}{1-\lambda}\left[1+ \lambda \rc \left(\frac{(n-2)\rc}{1+\rc (n-1)}\right)^{n-2}\right],
\end{equation}
which satisfy $\xi < 1$ and $\varsigma < \tau$, our result can be stated as follows.
For $\lambda \geq \xi$ the evolutionary dynamics depends on how the cost-to-benefit ratio $\cc/\bb$ compares to $1$ and to $\varsigma$.
If $\cc/\bb \leq 1$ (low costs), cooperation dominates defection, while if $\cc/\bb \geq \varsigma$ (high costs), defection dominates cooperation.
If $1 < \cc/\bb < \varsigma$ (intermediate costs), the dynamics are bistable.
For $\lambda < \xi$, the classification of possible evolutionary dynamics is as in the case $\lambda \geq \xi$, except that, if $\varsigma < \cc/\bb < \tau$, the dynamics are characterized by bistable coexistence, with $z= 0$ stable, $z_\textrm{L} \in (0,\hat{z})$ unstable, $z_\textrm{R} \in (\hat{z},1)$ stable, and $z=1$ unstable, where
\begin{equation}
\label{eq:hatzexpressers}
 \hat{z} = \frac{1+\rc}{[1+\rc (n-1)](1-\lambda)} .
\end{equation}

For $\rc \geq 0$, the critical values $\xi$, $\varsigma$, and $\tau$ are all increasing functions of $\rc$.
Hence, as relatedness $\rc$ increases, the regions of the parameter space where some level of expression of helping is stable expand at the expense of the region of dominant nonexpression.
Moreover, inside these regions the stable positive probability of expressing helping increases with $\rc$ (fig. \ref{fig:resultsfourkinds}.\emph{c}).
When synergies of scale are ``sufficiently'' negative ($\lambda < \xi$) and for intermediate cost-to-benefit ratios ($\varsigma < \cc/\bb < \tau$) relatedness and synergies interact in a nontrivial way, leading to saddle-node bifurcations as $\rc$ increases (fig. \ref{fig:resultsfourkinds}.\emph{c}).

\subsection{Connections with previous models}

\label{subsec:connections}

Our model without synergies of scale, for which the $\Gg(z)$ is linear in $z$ (section \ref{subsec:linear}) extends classical two-player matrix games between relatives (e.g. \citealp{Grafen1979}, \citealp[ch. 5-6]{Frank1998}) to the more general case of $n$-player linear games between relatives.
Indeed, for $n=2$, identifying scaled relatedness $\kappa$ with relatedness $r$, and up to normalization of the payoff matrices, equation \eqref{eq:ESSlinear} recovers \citet[eq. 9]{Grafen1979} and \citet[eq. 5.6]{Frank1998}.
Interestingly, \citet[p. 98]{Frank1998} considers a two-player model of helping with two pure strategies (``nesting'' or expressing a queen phenotype, and ``helping'' or expressing a sterile worker phenotype), which is a particular case of our model of nonexpresser-only traits.

Our results on whole-group traits with geometric returns (section \ref{subsec:whole} and app. \ref{sec:wholeapp}) extend the model studied by \citet[p.~198]{Hauert2006a} from the particular case of interactions between unrelated individuals ($\rc = 0$) to the more general case of interactions between relatives ($\rc \in [-1,1]$) and recover the result by \citet[p.~476]{Archetti2009} in the limit $\lambda \rightarrow 0$, in which the game is also called a ``volunteer's dilemma'' \citep{Diekmann1985}.
Although we restricted our attention to the cases of constant, decreasing, and increasing incremental benefits, it is clear that equation~\eqref{eq:Gzwhole2} applies to benefits $\bb_j$ of any shape.
Hence, general results about the stability of equilibria in public goods games \citep{Pena2014} with sigmoid benefits \citep{Bach2006,Archetti2011} carry over to games between relatives.

For their model of ``self-destructive cooperation'' in bacteria, \cite{Ackermann2008} assumed a nonexpresser-only trait with no synergies of scale, and a haystack model of population structure implying $\rc = (\No-N)/(\No(N-1))$, where $n = \No \geq N$ is the number of offspring among which the game is played (see eq. \eqref{eq:kappaackermann}).
Identifying our $\cc$ and $\bb$ with (respectively) their $\beta$ with $b$, the main result of \cite{Ackermann2008} (eq. 7 in their supplementary material) is recovered as a particular case of our result that the unique convergent stable strategy for this case is given by $z^* = [\rc (n-1) \bb-\cc]/[(1+\rc)(n-1) \bb]$ (eq. \eqref{eq:ESSlinear}).
The fact that in this example $\rc$ is a probability of coalescence within groups
shows that social interactions effectively occur between family members, and hence that kin selection is crucial to the understanding of self-destructive cooperation \citep{Gardner2008}.

As mentioned before, the analysis of other-only traits follows closely that of whole-group traits (see app. \ref{sec:otherapp}).
The model of altruistic helping in \cite{Eshel1988} considers such an other-only trait.
In their model, one individual in the group needs help, which can be provided (action A) or denied (action B) by its $n-1$ neighbors: a situation \citeauthor{Eshel1988} call the ``three brothers' problem'' when $n=3$.
Suppose that the cost for each helper is a constant $\varepsilon > 0$ independent on the number of expressers (\citet{Eshel1988}'s ``risk for each volunteer'', denoted by $c$ in their paper) and that the benefit for the individual in need when $k$ co-players offer help is given by $v_k$ (\citet{Eshel1988}'s ``gain function'', denoted by $b_k$ in their paper).
Then, if individuals need help at random, the payoffs for helping (A) and not helping (B) are given by $a_k = - \varepsilon (n-1)/n + v_k/n$ and $b_k = v_k/n$.
Defining $\cc = \varepsilon (n-1)/n$ and $\bb_k = v_k/(n-1)$, we have $a_k = - \cc + \bb_k$ and $b_k = \bb_k$.
Comparing these with the payoffs for whole-group traits in table \ref{table:threekinds}, it is apparent that the key difference between other-only traits and whole-group traits is that an expresser is not among the recipients of its own helping behavior.
As we show in appendix \ref{sec:otherapp}, our results for whole-group traits carry over to such other-only traits.
In particular, our results for whole-group traits with geometric benefits can be used to recover results 1,2, and 3 of \cite{Eshel1988} and to extend them from family-structured to spatially-structured populations.

Finally, \cite{VanCleve2013} discuss an $n$-player coordination game.
They assume payoffs given by $a_k = 1 + S (R/S)^{k/(n-1)}$ and $b_k = 1 + P (T/P)^{k/(n-1)}$, for positive $R,S,T$, and $P$, satisfying $R>T$, $P>S$ and $P>T$.
It is easy to see that both the direct effect $-\Cc(z)$ and the indirect effect $\Bb(z)$ are strictly increasing functions of $z$ having exactly one sign change.
This implies that, for $\rc \geq 0$, the evolutionary dynamics are characterized by bistability, with the basins of attraction of the two equilibria $z=0$ and $z=1$ being divided by the interior unstable equilibrium $z^*$.
Importantly, and in contrast to the social traits analyzed in this article, expressing the payoff dominant action A does not always qualify as a helping trait, as $\Bb(z)$ is negative for some interval $z \in [0,\hat{z})$.
As a result, increasing scaled relatedness $\rc$ can have mixed effects on the location of $z^*$.
Both of these predictions are well supported by the numerical results reported by \cite{VanCleve2013}, where increasing $\rc$ leads to a steady increase in $z^*$ for $R=2$, $S=0.5$, $P=1.5$, $T=0.25$, and a steady decrease in $z^*$ for $R=2$, $S=0.5$, $P=1.5$, $T=1.25$, see their figure 5.
This illustrates that relatedness (and thus spatial structure) plays an important role not only in the specific context of helping games but also in the more general context of nonlinear multiplayer games.

\section{Discussion}

We have shown that, when phenotypic differences are small, the selection gradient on a mixed strategy of a symmetric two-strategy $n$-player matrix game is proportional to the average inclusive payoff gain to an individual switching strategies, and that this can be written as a polynomial in Bernstein form (eq. \eqref{eq:Gzbinom}).
As a result, convergence stability of strategies in spatially structured populations can be determined from the shape of the inclusive gain sequence (eq.~\eqref{eq:inclusivegains}) and the mathematical properties of polynomials in Bernstein form \citep{Farouki2012,Pena2014}.
We applied these results to the evolution of helping under synergies of scale and kind, and unified and extended previous analysis.
The most important conclusion we reach is that, although an increase in (scaled) relatedness $\rc$ always tempers the social dilemma faced by cooperative individuals in a helping game, how the social dilemma is relaxed crucially depends on the synergies of kind
and scale involved.

\begin{figure}[t]
\centering
\includegraphics[width=\textwidth]{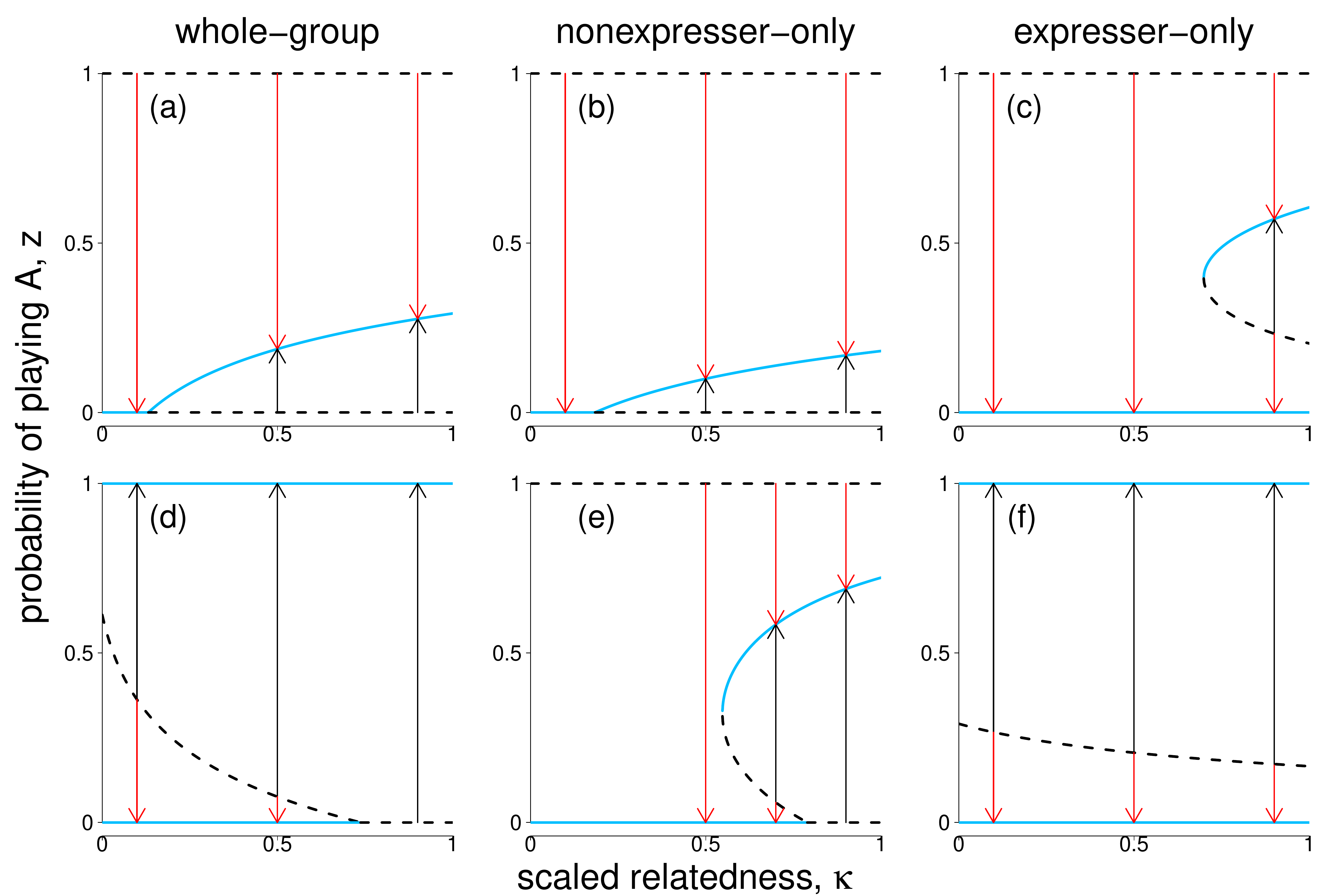}
\caption{Bifurcation diagrams for whole-group (\emph{a}, \emph{d}), nonexpresser-only (\emph{b}, \emph{e}), and expresser-only (\emph{c}, \emph{f}) traits with geometric benefits.
The scaled relatedness coefficient $\rc \ge 0$ serves as a control parameter.
Arrows indicate the direction of evolution for the probability of expressing helping.
Solid lines stand for stable equilibria; dashed lines for unstable equilibria.
\emph{a}, \emph{b}, \emph{c}, Negative synergies of scale ($\lambda=0.7$) and low cost-to-benefit ratio ($\cc/\bb=3.5$).
\emph{d}, \emph{e}, \emph{f}, Positive synergies of scale ($\lambda=1.25$) and high cost-to-benefit ratio ($\cc/\bb=15$).
In all plots, $n=20$.
}
\label{fig:resultsfourkinds}
\end{figure}

The simplest case is the one of whole-group traits (fig.~\ref{fig:threekinds}\emph{a}).
Since there are no synergies of kind, only synergies of scale can introduce frequency dependent selection.
For $\rc \geq 0$, negative (resp. positive) synergies of scale induce negative (resp. positive) frequency-dependent selection.
Moreover, increasing relatedness can transform a game in which defection is dominant (Prisoner's Dilemma) into a game in which cooperation and defection coexist (Snowdrift or anti-coordination game) when synergies of scale are negative (fig. \ref{fig:resultsfourkinds}.\emph{a}), or into a game in which both cooperation and defection are stable (Stag Hunt or coordination game) when synergies of scale are positive (fig. \ref{fig:resultsfourkinds}.\emph{d}).

More complex interactions between relatedness and frequency dependence can arise when there are both synergies of kind and scale.
For nonexpresser-only traits (fig.~\ref{fig:threekinds}.\emph{b}), synergies of kind are negative and helping is altruistic, so that in the absence of relatedness defection dominates cooperation (as in a Prisoner's Dilemma).
When synergies of scale are absent (linear benefits) or negative (diminishing incremental benefits), both the direct and the indirect effect are decreasing in $z$ and selection is negative frequency-dependent.
In this case, increasing relatedness might turn the game into a Snowdrift or an anti-coordination game,
where the probability of cooperating is an increasing function of relatedness (fig. \ref{fig:resultsfourkinds}.\emph{b}).
Contrastingly, when synergies of scale are positive (increasing incremental benefits) the indirect effect may become unimodal in $z$.
This paves the way for new patterns of evolutionary dynamics and bifurcations.
For the particular case of geometric benefits, we find that there is a range of cost-to-benefit ratios such that, for sufficiently strong positive synergies of scale, increasing relatedness induces a saddle-node bifurcation whereby two internal equilibria appear, the leftmost unstable and the rightmost stable (fig. \ref{fig:resultsfourkinds}.\emph{e}).
After the bifurcation occurs, the evolutionary dynamics are characterized by bistable coexistence, where the first stable equilibrium is pure defection ($z=0$) and the second is a mixed equilibrium in which individuals help with a positive probability ($z_\textrm{R}$).

For expresser-only traits (where synergies of kind are positive, fig.~\ref{fig:threekinds}\emph{c}) a similar interaction between relatedness and synergies occurs.
When synergies of scale are absent or positive, $\Gg(z)$ is increasing in $z$ for $\rc \ge 0$.
In this case, increasing relatedness might turn a scenario reminiscent of the Prisoner's Dilemma into a Stag Hunt or coordination game, where the size of the basin of attraction of the cooperative equilibrium is an increasing function of relatedness (fig. \ref{fig:resultsfourkinds}.\emph{f}).
Contrastingly, if synergies of scale are negative, relatedness may interact nontrivially with synergies to produce a dynamical outcome which is qualitatively identical to that arising from nonexpresser-only traits with positive synergies of scale, namely, bistable coexistence (fig. \ref{fig:resultsfourkinds}.\emph{c}).

The three kinds of helping traits we considered are also different in the conditions they impose on the origin and the maintenance of helping.
To see this, consider a payoff cost $\cc$ so large that the direct sequence $\dk$ is negative.
For the case of unrelated individuals ($\rc = 0$) this implies that B dominates A so that $z=0$ is the only stable strategy.
We ask what happens when $\rc$ is increased and focus on the stability of the end-points $z=0$ and $z=1$.

For whole-group traits, the indirect gains from switching when co-players are all defectors ($e_0$) and when co-players are all helpers ($e_{n-1}$) are both positive.
This opens up the opportunity for both (i) $z=0$ to be destabilized if $\rc > -d_0/e_0$, and (ii) $z=1$ to be stabilized if $\rc > -d_{n-1}/e_{n-1}$, which underlies the classical effect that increasing relatedness can destabilize defection and stabilize helping.

In contrast, one of these two scenarios is missing for nonexpresser-only and expresser-only traits.
For nonexpresser-only traits, we have $e_0 > 0$ and $e_{n-1} = 0$ irrespectively of the shape of the benefit sequence.
Hence, although $z=0$ can be destabilized by increasing $\rc$ (allowing for some level of helping to be evolutionarily accessible from $z=0$), $z=1$ can never be stabilized and so full helping is never an evolutionary (convergent) stable point.
Exactly the opposite happens for expresser-only traits, where $e_n > 0$ but $e_0 = 0$.
As a result, $z=1$ can become stable (if $\rc > -d_{n-1}/e_{n-1}$) but $z=0$ can never be destabilized by increasing $\rc$.
This implies that, under our assumptions, an expresser-only trait with high costs ($\cc > \bb_n$) can never evolve from a monomorphic population of nonexpressers ($z=0$),
and this for any value of $\rc$.

The kind of social trait also has a big impact on the amount of (scaled) relatedness required to make stable some level of helping.
This quantitative effect is illustrated in figure \ref{fig:resultsfourkinds}.
When synergies of scale are negative ($\lambda=0.7$) and the cost-to benefit ratio is relatively low ($\cc/\bb=3.5$), for whole-group and nonexpresser-only traits, moderate amounts of relatedness ($\rc \approx 0.132$ for whole-group traits, $\rc \approx 0.185$ for nonexpresser-only traits) are sufficient for a non-zero level of expression of helping to be stable (fig. \ref{fig:resultsfourkinds}.\emph{a} and \ref{fig:resultsfourkinds}.\emph{b}).
In contrast, a comparatively large amount of relatedness ($\rc \approx 0.699$) is required for some non-zero level of helping to be stable if the trait is expresser-only (fig. \ref{fig:resultsfourkinds}.\emph{c}).
In the case of positive synergies of scale ($\lambda=1.25$) and relatively high cost-to-benefit ratio ($\cc/\bb=15$), full expression of helping is stable already with $\rc = 0$
for whole-group and expresser-only traits (fig. \ref{fig:resultsfourkinds}.\emph{d} and \ref{fig:resultsfourkinds}.\emph{f}).
Contrastingly, for nonexpresser-only traits, a positive probability of expressing helping is stable only for large values of relatedness ($\rc > \rc^* \approx 0.548$, fig. \ref{fig:resultsfourkinds}.\emph{e}).

We modeled social interactions by assuming that actions implemented by players are discrete.
This is in contrast to many kin-selection models of games between relatives, which assume a continuum of pure actions in the form of continuous amounts of effort devoted to some social activity (e.g., \citealt{Frank1994,Johnstone1999,Reuter2001,Wenseleers2010}).
Such continuous-action models have the advantage that the ``fitness function'' or ``payoff function'' (the counterpart to our eq. \eqref{eq:piapprox}) usually takes a simple form that facilitates mathematical analysis. 
On the other hand, there are situations where individuals can express only a few behavioral alternatives or morphs, such as worker and queen in the eusocial Hymenoptera \citep{Wheeler1986}, different behavioral tactics in foraging (e.g., ``producers'' and ``scroungers'' in house sparrows \emph{Passer domesticus}; \citealp{Barnard1981})
and hunting (e.g., lionesses positioned as ``wings'' and others positioned as ``centres'' in collective hunts; \citealp{Stander1992}), or distinct phenotypic states (e.g., capsulated and non-capsulated cells in \emph{Pseudomonas fluorescens}; \citealp{Beaumont2009}).
These situations are more conveniently modeled by means of a discrete-action model like the one presented here, but we expect that our qualitative results about the interaction between synergy and relatedness carry over to continuous-action models.

Synergistic interactions are likely to be much more common in nature than additive interactions where both synergies of scale and kind are absent.
Given the local demographic structure of biological populations,
interactions between relatives are also likely to be the rule rather than the exception.
Empirical work should thus aim at measuring not only the genetic relatedness of interactants and the fitness costs and benefits of particular actions,
but also at identifying the occurrences of positive and negative synergies of kind and scale,
as it is the interaction between synergies and relatedness which determines the qualitative outcomes of the evolutionary dynamics of helping (fig. \ref{fig:resultsfourkinds}).

\section{Acknowledgements}

This work was partly supported by Swiss NSF Grants PBLAP3-145860 (to JP) and PP00P3-123344 (to LL).

\newpage

\appendix

\setcounter{equation}{0}
\renewcommand{\theequation}{A.\arabic{equation}}

\section{The haystack model}
\label{sec:kappaapp}

Many models of social interactions have assumed different versions of the haystack model (e.g., \citealp{Matessi1976,Ackermann2008}), where several rounds of unregulated reproduction can occur within groups before a round of complete dispersal \citep{MaynardSmith1964} so that competition is effectively global.
In these cases, as we will see below, $\rc$ takes the simpler interpretation of the coalescence probability of the gene lineage of two interacting individuals in their group.
Here, we calculate $\rc$ for different variants of the haystack model.

The haystack model can be seen as a special case of the island model where dispersal is complete and where dispersing progeny compete globally.
In this context, the fecundity of an adult is the number of its offspring reaching the stage of global density-dependent competition.
The conception of offspring may occur in a single or over multiple rounds of reproduction, so that a growth phase within patches is possible.
In this context, the number $N$ of  ``adults'' is better thought of as the number of founding individuals (or lineages, or seeds) on a patch.

Two cases need to be distinguished when it comes to social interactions.
First, the game can be played between the adult individuals (founders) in which case
\begin{equation}
\rc=0,
\end{equation}
since relatedness is zero among founders on a patch and there is no local competition.
Alternatively, the game is played between offspring after reproduction and right before their dispersal.
In this case two individuals can be related since they can descend from the same founder.
Since there is no local competition, $\rc$ is directly the relatedness between two interacting offspring and is obtained as the probability that the two ancestral lineages of two randomly sampled offspring coalesce in the same founding individual (relatedness in the island model is defined as the cumulative coalescence probability over several generations,
see e.g., \citealp{Rousset2004}, but owing to complete dispersal gene lineages can only coalesce in founders).

In order to evaluate $\rc$ for the second case, we assume that, after growth, exactly $\No$ offspring are produced and that the game is played between them ($n=\No$).
Founding individuals, however, may contribute a variable number of offspring.
Let us denote by $\J_i$ the random number of offspring descending from the ``adult'' individual $i=1,2,...,N$ on a representative patch after reproduction, i.e., $\J_i$ is the size of lineage $i$.
Owing to our assumption that the total number of offspring is fixed, we have $\No=\J_{1}+\J_{2}+...+\J_{N}$, where the $\J_i$'s are exchangeable random variables (i.e., neutral process, $\delta=0$).
The coalescence probability $\rc$ can then be computed as the expectation of the ratio of the total number of ways of sampling two offspring from the same founding parent to the total number of ways of sampling two offspring:
\begin{equation}
 \rc=\Esp\left[ \sum_{i=1}^{N} \frac{\J_{i}(\J_{i}-1)}{\No(\No-1)}\right]= N\left(\frac{\sigma^2+\mu^2-\mu}{\No(\No-1)} \right),
\end{equation}
where the second equality follows from exchangeability, $\mu=\Esp\left[\J_i\right]$ is the expected number of offspring descending from any individual $i$, and $\sigma^2=\Esp\left[(\J_i-\mu)^2\right]$ is the corresponding variance.
Due to the fact that the total number of offspring is fixed, we also necessarily have $\mu=\No/N$ (i.e., $\No=\Esp\left[\No\right]=\Esp\left[\J_{1}+\J_{2}+...+\J_{N}\right]=N\mu$), whereby
\begin{equation}
 \label{eq:haystackkappa}
 \rc=\frac{\No-N}{N(\No-1)}+\frac{\sigma^2N}{\No(\No-1)},
\end{equation}
which holds for any neutral growth process.

We now consider different cases:

(i) Suppose that there is no variation in offspring production between founding individuals, as in the life cycle described by \citet{Ackermann2008}.
Then $\sigma^2=0$, and equation \eqref{eq:haystackkappa} simplifies to
\begin{equation}
 \label{eq:kappaackermann} 
 \rc=\frac{(\No-N)}{N(\No-1)} .
\end{equation}

(ii) Suppose that each of the $\No$ offspring has an equal chance of descending from any founding individual, so that each offspring is the result of a sampling event (with replacement) from a parent among the $N$ founding individuals.
Then, the offspring number distribution is binomial with parameters $\No$ and $1/N$, whereby $\sigma^2=(1-1/N)\No/N$.
Substituting into equation~\eqref{eq:haystackkappa} produces
\begin{equation}
\rc=\frac{1}{N}.
\end{equation}
In more biological terms, this case results from a situation where individuals produce offspring according to a Poisson process and where exactly $\No$ individuals are kept for interactions (i.e., the conditional branching process of population genetics; \citealp{Ewens04}).

(iii) Suppose that the offspring distribution follows a beta-binomial distribution,
with number of trials $\No$ and shape parameters $\alpha>0$ and $\beta=\alpha(N-1)$.
Then, $\mu=\No/N$ and
\[
\sigma^2=\frac{\No(N-1)(\alpha N+\No)}{N^2(1+\alpha N)},
\]
which yields
\begin{equation}
\rc=\frac{1+\alpha}{1+\alpha N}.
\end{equation}
In more biological terms, this reproductive scheme results from a situation where individuals produce offspring according to a negative binomial distribution (larger variance than Poisson, which is recovered when $\alpha \to \infty$), and where exactly $\No$ individuals are kept for interactions.

\setcounter{equation}{0}
\renewcommand{\theequation}{B.\arabic{equation}}

\section{Gains from switching and the gain function}
\label{sec:gzapp}

In the following we establish the expressions for $\Cc(z)$ and $\Bb(z)$ given in equations \eqref{eq:Cz}--\eqref{eq:Bz}; equation \eqref{eq:Gzbinom} is then immediate from the definition of $f_k$ \eqref{eq:inclusivegains} and the identity $\Gg(z) = - \Cc(z) + \rc \Bb(z)$.

Recalling the definitions of $\Cc(z)$ and $\Bb(z)$ from equation~\eqref{eq:Gz} as well as the definitions of $d_k$ and $e_k$ from equations \eqref{eq:directgains}--\eqref{eq:indirectgains} we need to show
\begin{align}
\label{ap-b-1} \frac{\partial \pi(z_\focal,z_\deme)}{\partial z_\focal} \bigg|_{z_\focal=z_\deme=z} & = \sum_{k=0}^{n-1} {n-1 \choose k} z^k (1-z)^{n-1-k} \left[a_k - b_k\right],    \\
\label{ap-b-2} \frac{\partial \pi(z_\focal,z_\deme)}{\partial z_\deme} \bigg|_{z_\focal=z_\deme=z}  & = \sum_{k=0}^{n-1} {n-1 \choose k} z^k (1-z)^{n-1-k} \left[k \Delta a_{k-1} + (n-1-k) \Delta b_k \right],
\end{align}
where the function $\pi$ has been defined in equation \eqref{eq:piapprox}.
Equation \eqref{ap-b-1} follows directly by taking the partial derivative of $\pi$ with respect to $z_\bullet$ and evaluating at $z_\bullet = z_\circ = z$, so it remains to establish equation \eqref{ap-b-2}.

Our derivation of equation \eqref{ap-b-2} uses properties of polynomials in Bernstein form \citep{Farouki2012}.
Such polynomials, which in general can be written as $\sum_{k=0}^{m} \binom{m}{k} x^k (1-x)^{m-k} c_k$, where $x \in [0,1]$, satisfy
\begin{equation*}
 \label{eq:derivative}
 \frac{\dd}{\dd x}\sum_{k=0}^{m} {m \choose k} x^k (1-x)^{m-k} c_k = m \sum_{k=0}^{m-1} \binom{m-1}{k} x^k (1-x)^{m-1-k} \Delta c_k.
\end{equation*}

Applying this property to equation \eqref{eq:piapprox} and evaluating the resulting partial derivative at $z_\bullet = z_\deme = z$, yields
\begin{equation}\label{eq:use-later}
\frac{\partial \pi(z_\focal,z_\deme)}{\partial z_\deme} \Bigg|_{z_\focal=z_\deme=z} = (n-1) z \sum_{k=0}^{n-2} \binom{n-2}{k} z^k (1-z)^{n-2-k} \Delta a_k + (n-1)(1-z) \sum_{k=0}^{n-2} \binom{n-2}{k} z^k (1-z)^{n-2-k} \Delta b_k.
\end{equation}

In order to obtain equation \eqref{ap-b-2} from equation \eqref{eq:use-later} it then suffices to establish
\begin{equation}
\label{eq:multx}
 x \sum_{k=0}^{m-1} \binom{m-1}{k} x^k (1-x)^{m-1-k} c_k = \sum_{k=0}^{m} \binom{m}{k} x^k (1-x)^{m-k} \frac{k c_{k-1}}{m}
\end{equation}
and
\begin{equation}
\label{eq:mult1-x}
 (1-x)\sum_{k=0}^{m-1} \binom{m-1}{k} x^k (1-x)^{m-1-k} c_k = \sum_{k=0}^{m} \binom{m}{k} x^k (1-x)^{m-k} \frac{(m-k)c_k}{m},
\end{equation}
as applying these identities to the terms on the right side of equation \eqref{eq:use-later} yields the right side of equation \eqref{ap-b-2}.

Let us prove equation \eqref{eq:multx} (eq. \eqref{eq:mult1-x} is proven in a similar way).
Starting from the left side of equation \eqref{eq:multx}, we multiply and divide by $m/(k+1)$ and distribute $x$ to obtain
\[
 x \sum_{k=0}^{m-1} \binom{m-1}{k} x^k (1-x)^{m-1-k} c_k = \sum_{k=0}^{m-1} \frac{m}{k+1} \binom{m-1}{k} x^{k+1} (1-x)^{m-(k+1)} c_k \frac{k+1}{m} .
\]
Applying the identity $\binom{r}{k}=\frac{r}{k}\binom{r-1}{k-1}$ and changing the index of summation to $k=k+1$, we get
\[
 x \sum_{k=0}^{m-1} \binom{m-1}{k} x^k (1-x)^{m-1-k} c_k = \sum_{k=1}^{m} \binom{m}{k} x^k (1-x)^{m-k} \frac{k c_{k-1}}{m}.
\]
Finally, changing the lower index of the sum by noting that the summand is zero when $k=0$ gives equation \eqref{eq:multx}.

\setcounter{equation}{0}
\renewcommand{\theequation}{D.\arabic{equation}}

\setcounter{equation}{0}
\renewcommand{\theequation}{C.\arabic{equation}}

\section{Whole-group traits with geometric benefits}
\label{sec:wholeapp}

With geometric benefits, we have $\Delta \bb_k = \bb \lambda^k$, so that the inclusive gains from switching for whole-group traits are given by $f_k = - \cc + \left[1+\rc (n-1)\right] \bb \lambda^{k}$.
Using the formula for the probability generating function of a binomial random variable, equation~\eqref{eq:Gzbinom} can be written as
\begin{equation}
\label{eq:Sgeowh}
 \Gg(z) = -\cc + \left[1+\rc (n-1)\right] \bb \left(1-z+\lambda z\right)^{n-1}.
\end{equation}
As $\Gg(z)$ is either decreasing ($\lambda < 1$) or increasing ($\lambda > 1$) in $z$, A (resp. B) is a dominant strategy if and only if $\min \left[\Gg(0), \Gg(1)\right] \ge 0$ (resp. if and only if $\max \left[\Gg(0), \Gg(1)\right] \le 0$).
Using equation~\eqref{eq:Sgeowh} to calculate $\Gg(0)$ and $\Gg(1)$ then yields the critical cost-to-benefit ratios $\varepsilon = \min \left[\Gg(0), \Gg(1)\right]$ and $\vartheta = \max \left[\Gg(0), \Gg(1)\right]$ given in equation~\eqref{eq:epsilonwhole}.
The value of $z^*$ given in equation~\eqref{eq:zwholegroup} is obtained by solving $\Gg(z^*) = 0$.

\setcounter{equation}{0}
\renewcommand{\theequation}{D.\arabic{equation}}

\section{Other-only traits}
\label{sec:otherapp}

In contrast to what happens in whole-group traits, individuals expressing an other-only trait are automatically excluded from the consumption of the good they create,
although they can still reap the benefits of goods created by other expressers in their group.
Payoffs for such other-only traits are given by $a_k = - \gamma + \bb_k$ and $b_k = \bb_k$, so that the inclusive gains from switching are given by $f_k = - \gamma + \kappa \left[ k \Delta \bb_{k-1} + (n-1-k) \Delta \bb_k \right]$.
For this payoff constellation, it is straightforward to obtain the indirect benefits $\Bb(z)$ from equation \eqref{eq:use-later} in appendix \ref{sec:gzapp}.
Observing that $\Delta a_k = \Delta b_k = \Delta \bb_k$ holds for all $k$, we have
\begin{equation*}
\Bb(z) = \frac{\partial \pi(z_\focal,z_\deme)}{\partial z_\deme} \Bigg|_{z_\focal=z_\deme=z} = \sum_{k=0}^{n-2} {n-2 \choose k} z^k (1-z)^{n-2-k} (n-1) \Delta \bb_k .
\end{equation*}
Using equation \eqref{eq:Cz} and the equality $a_k - b_k = - \cc$, we have that the direct benefit is given by $- \Cc(z) = - \cc$.
Substituting these expressions for $\Cc(z)$ and $\Bb(z)$ into equation \eqref{eq:Gz}, we obtain
\begin{equation}
\label{eq:gzother}
 \Gg(z) = \sum_{k=0}^{n-2} \binom{n-2}{k} z^k (1-z)^{n-2-k} \left[- \cc + \rc (n-1) \Delta \bb_k \right] .
\end{equation}

If $\rc \leq 0$, our assumption that the benefit sequence is increasing implies
that $\Gg(z)$ is always negative, so that $z=0$ is the only stable point (defection dominates cooperation).

To analyze the case where $\rc > 0$, it is convenient to observe that equation \eqref{eq:gzother} is of a similar form as equation \eqref{eq:Gzwhole}.
The only differences are that the summation in equation \eqref{eq:gzother} extends from $0$ to $n-2$ (rather than $n-1$) and that the term multiplying the incremental benefit $\Delta \bb_k$ is given by $\rc (n-1)$ (rather than $1 + \rc (n-1)$).
All the results obtained for whole-group traits can thus be easily translated to the case of other-only traits.

Specifically, we have the following characterization of the resulting evolutionary dynamics.
In the absence of synergies of scale, selection is frequency-independent with defection dominating cooperation if $\rc < \gamma/[(n-1) \beta]$ and cooperation dominating defection if $\rc > \gamma/[(n-1) \beta]$.
With negative synergies of scale, the gain function is decreasing in $z$ (negative frequency dependence).
Defection dominates cooperation (so that $z=0$ is the only convergent stable strategy) if $\cc \ge \rc (n-1) \Delta \bb_0$, whereas cooperation dominates defection if $\cc \le \rc (n-1) \Delta \bb_{n-2}$ holds.
If $\rc (n-1)  \Delta \bb_{n-2} < \cc < \rc (n-1) \Delta \bb_0$ holds, the unique convergent stable strategy $z^*$ features coexistence.
With positive synergies of scale, the gain function is increasing in $z$ (positive frequency dependence).
Defection dominates cooperation if $\cc \ge \rc (n-1) \Delta \bb_{n-2}$, whereas cooperation dominates defection if $\cc \le \rc (n-1) \Delta \bb_0$.
If $\rc (n-1) \Delta \bb_0 < \cc < \rc (n-1)\Delta \bb_{n-2}$, there is bistability: both $z=0$ and $z=1$ are stable and there is a unique, unstable interior point $z^*$ separating the basins of attraction of these two stable strategies.

When benefits are geometric \eqref{eq:geombenefits}, the gain function is given by
\[
 \Gg(z) = - \gamma + \kappa(n-1) \bb (1-z+\lambda z)^{n-2} ,
\]
so that, for $\lambda \neq 1$, the evolutionary dynamics are similar to the case of whole-group traits after redefining the critical cost-to-benefit ratios as
\begin{equation*}
 \label{eq:epsilonother}
 \varepsilon = \text{min}\left(\rc (n-1), \lambda^{n-2} \rc (n-1) \right) \qquad\text{and}\qquad  \vartheta = \text{max}\left(\rc (n-1), \lambda^{n-2} \rc (n-1) \right)
\end{equation*}
and letting
\begin{equation*}
\label{eq:zother}
z^*= \frac{1}{1-\lambda} \left[ 1-\left(\frac{\cc}{\bb \rc (n-1)}\right)^{\frac{1}{n-2}} \right] .
\end{equation*}

\setcounter{equation}{0}
\renewcommand{\theequation}{E.\arabic{equation}}

\section{Nonexpresser-only traits}
\label{sec:nonexpressersapp}

For nonexpresser-only traits, the inclusive gains from switching are given by
\begin{equation}
 \label{eq:fknonexpressers}
 f_k = - \cc - \bb_k + \rc (n-1-k) \Delta \bb_k.
\end{equation}

\subsection{Negative synergies of scale}
\label{subsec:nonexpressersappconcave}

When synergies of scale are negative, we have the following general result.

\begin{result}[Nonexpresser-only traits with negative synergies of scale]
\label{res:nonexpressersconcave}
Let $\fk$ be given by equation \eqref{eq:fknonexpressers} with $\bb_0 = 0$, $\bb_k$ increasing and $\Delta \bb_k$ decreasing in $k$ and let
$\rc \geq 0$ (the case $\rc < 0$ is trivial).
Then
\begin{enumerate}
\item If $\cc \ge \rc (n-1) \Delta \bb_0$, $z=0$ is the only stable point (B dominates A).
\item If $\cc < \rc (n-1) \Delta \bb_0$, both $z=0$ and $z=1$ are unstable and there is a unique internal stable point $z^* \in (0,1)$ (coexistence).
\end{enumerate}
\end{result}

To prove this result, we start by observing that the assumptions in the statement imply that $f_k$ is decreasing in $k$.
In particular, we have $f_{n-1} < f_0$.
Consequently, if $f_0 \le 0$ (which holds if and only if $\cc \ge \rc (n-1) \Delta \bb_0$) the inclusive gain sequence has no sign changes and its initial sign is negative.
Observing that $f_{n-1} = -\cc - \bb_{n-1} < 0$ always holds true, the inequality $f_0 > 0$ (which holds if and only if $\cc < \rc (n-1) \Delta \bb_0$) implies that the decreasing sequence $\fk$ has one sign change and that its initial sign is positive.
Result \ref{res:nonexpressersconcave} is then obtained by an application of \citet[result 3]{Pena2014}.

\subsection{Geometric benefits}
\label{subsec:nonexpressersappgeometric}

For geometric benefits, we obtain the following result.

\begin{result}[Nonexpresser-only traits with geometric benefits]
\label{res:nonexpressersgeom}
Let $\fk$ be given by equation \eqref{eq:fknonexpressers} with $\bb_k$ given by equation \eqref{eq:geombenefits} and let $\rc \geq 0$ and $n>2$ (the cases $\rc < 0$ or $n=2$ are trivial).
Moreover, let $\varrho$, $\zeta$ and $\eta$ be defined by equations \eqref{eq:varrho} and \eqref{eq:zeta}.
Then
\begin{enumerate}
\item If $\lambda \leq \varrho$, $\Gg(z)$ is nonincreasing in $z$. Furthermore:
\begin{enumerate}
\item If $\cc/\bb < \zeta$, both $z=0$ and $z=1$ are unstable and there is a unique internal stable point $z^* \in (0,1)$ (coexistence).
\item If $\cc/\bb \ge \zeta$, $z=0$ is the only stable point (B dominates A).
\end{enumerate}
\item If $\lambda > \varrho$, $\Gg(z)$ is unimodal in $z$ with mode given by $\hat{z}=\frac{\rc \left[(n-2)\lambda - (n-1)\right]-1}{[1+\rc (n-1)](\lambda-1)}$. Furthermore:
\begin{enumerate}
 \item If $\cc/\bb \le \zeta$, both $z=0$ and $z=1$ are unstable and there is a unique internal stable point $z^* > \hat{z}$ (coexistence).
 \item If $\zeta < \cc/\bb < \eta$, there are two interior singular points $z_\mathrm{L}$ and $z_\mathrm{R}$ satisfying $z_\mathrm{L} < \hat{z} < z_\mathrm{R}$.
The points $z=0$ and $z_\mathrm{R}$ are stable, whereas $z_\mathrm{L}$ and $z=1$ are unstable (bistable coexistence).
 \item If $\cc/\bb \ge \eta$, then $z=0$ is the only stable point (B dominates A).
\end{enumerate}
\end{enumerate}
\end{result}

Observing that $\varrho > 1$ holds for $\kappa \geq 0$ and that the case $\lambda = 1$ (no synergies of scale) is trivial, we can prove this result by considering three cases: (i) $\lambda < 1$, (ii)
$1 < \lambda \leq \varrho$, and (iii) $\varrho < \lambda$.

For $\lambda < 1$, we have negative synergies of scale and hence result \ref{res:nonexpressersconcave} applies with $\Delta \bb_0 = \bb$.
Recalling the definition of $\zeta = \rc (n-1)$ from equation \eqref{eq:zeta} and rearranging, this yields result \ref{res:nonexpressersgeom}.1 for the case $\lambda \leq 1 < \varrho$.

To obtain the result for the remaining two cases, we calculate the first and second forward differences of the benefit sequence \eqref{eq:geombenefits} and substitute them into
\begin{equation*}
 \Delta f_k = - (1+\rc) \Delta \bb_k + \rc (n-2-k) \Delta^2 \bb_k , \ k=0,1,\ldots,n-2 .
\end{equation*}
to obtain
\begin{equation*}
  \Delta f_k = \bb \lambda^k \left\{ \rc \left[(n-2)\lambda -(n-1) \right] -1 + \rc (1-\lambda) k \right\} , \ k=0,1,\ldots,n-2.
\end{equation*}
For $\lambda > 1$, the sequence $\Dfk$ is decreasing in $k$ and hence can have at most one sign change.
Moreover, since $\Delta f_{n-2}=-\bb\lambda^{n-2}(1+\rc) < 0$ always holds true, the sign pattern of $\Dfk$
depends exclusively on how $\Delta f_0 = \bb \left\{ \rc \left[(n-2)\lambda -(n-1) \right] -1 \right\}$ compares to zero. Observe, too, that $f_{n-1} < 0$ always holds true and that the sign of $f_0$ is identical to the sign of $\zeta - \cc/\bb$.

Consider the case $1 < \lambda \leq \varrho$.
Recalling the definition of $\varrho$ (eq. \eqref{eq:varrho}) we then have $\Delta f_0 \leq 0$, implying that $\Dfk$ has no sign changes and that its initial sign is negative, i.e., $\fk$ is nonincreasing.
Hence, if $f_0 \leq 0$ (which holds if and only if $\cc/\bb \geq \zeta$), the inclusive gain sequence has no sign changes and its initial sign is negative.
Otherwise, that is, if $\cc/\bb < \zeta$ holds, we have $f_0 > 0 > f_{n-1}$ so that the inclusive gain sequence has one sign change and its initial sign is positive.
Result \ref{res:nonexpressersgeom}.1 then follows from \citet[result 3]{Pena2014}.

For $\lambda > \varrho$ we have $\Delta f_0 > 0$, implying that $\Dfk$ has one sign change from $+$ to $-$, i.e., $\fk$ is unimodal.
This implies that the gain function $\Gg(z)$ is also unimodal with its mode $\hat{z}$ being determined by $\Gg'(\hat z) = 0$ \citep[section 3.4.3]{Pena2014}.
Using the assumption of geometric benefits, we can express $\Gg(z)$ is closed form as
 \begin{equation*}
 \Gg(z) = -\cc + \frac{\bb}{\lambda-1} + \bb \left\{\rc (n-1) - \frac{1}{\lambda-1} - \left[1+\rc (n-1)\right] z \right\} \left(1-z+\lambda z\right)^{n-2}
 \end{equation*}
with corresponding derivative
\begin{equation*}
 \Gg'(z) = (n-1) \bb (\lambda-1) \left(1-z+\lambda z\right)^{n-3} \left\{ \rc (n-2) - \frac{1+\rc}{\lambda-1} - \left[1+\rc (n-1)\right] z \right\}.
\end{equation*}
Solving $\Gg'(\hat z) = 0$ then yields $\hat z$ as given in equation \eqref{eq:hatznonexpressers}.
The corresponding maximal value of the gain function is
\begin{equation*}
\label{eq:zghatnonexpressers}
\Gg(\hat{z}) = -\cc+\frac{\bb}{\lambda-1}\left[ 1+ \rc \lambda \left( \frac{(n-2)\rc \lambda}{1+\rc (n-1)} \right)^{n-2} \right] .
\end{equation*}
Result \ref{res:nonexpressersgeom}.2 follows from an application of \citet[result 5]{Pena2014} upon noticing that $f_0 \ge 0$ (precluding the stability of $z = 0$ and ensuring $\Gg(\hat z) > 0$) holds if and only if $\cc/\bb \le \zeta$ and that $\Gg(\hat{z})\leq 0$ (ensuring that B dominates A) is satisfied if and only if $\cc/\bb \geq \eta$.
(We note that the range of cost-to-benefit ratios $\cc/\bb$ for which bistable coexistence occurs is nonempty, that is $\eta > \zeta$ holds.
Otherwise there would exist $\cc/\bb$ satisfying both $\cc/\bb \le \zeta$ and $\cc/\bb \ge \eta$ which in light of
result \ref{res:nonexpressersgeom}.2.(a) and result \ref{res:nonexpressersgeom}.2.(c) is clearly impossible.)

\setcounter{equation}{0}
\renewcommand{\theequation}{F.\arabic{equation}}

\section{Expresser-only traits}
\label{sec:expressersapp}

For expresser-only traits, the inclusive gains from switching are given by
\begin{equation}
 \label{eq:fkexpressers}
 f_k = - \cc + \bb_{k+1} + \rc k \Delta \bb_k .
\end{equation}

\subsection{Positive synergies of scale}
\label{subsec:expressersappconvex}

In the case of positive synergies of scale, we have the following general result.

\begin{result}[Expresser-only traits with positive synergies of scale]
\label{res:expressersconvex}
Let $\fk$ be given by equation \eqref{eq:fkexpressers} with $\bb_k$ and $\Delta \bb_k$ increasing in $k$ and let $\rc \geq 0$.
Then
\begin{enumerate}
\item If $\cc \le \bb_1$, $z=1$ is the only stable point (A dominates B).
\item If $\bb_1 < \cc < \bb_n+ \rc (n-1) \Delta \bb_{n-1}$, both $z=0$ and $z=1$ are stable and there is a unique internal unstable point $z^* \in (0,1)$ (bistability).
\item If $\cc \ge \bb_n + \rc (n-1) \Delta \bb_{n-1}$, $z=0$ is the only stable point (B dominates A).
\end{enumerate}
\end{result}

The arguments used for deriving this result are analogous to those used for deriving the results for the case of nonexpresser-only traits and negative synergies of scale
(result \ref{res:nonexpressersconcave} in app. \ref{sec:nonexpressersapp}).
The assumptions in the statement of the result imply that $f_k$ is increasing in $k$. In particular, we have $f_0 < f_{n-1}$.
The sign pattern of the inclusive gain sequence thus depends on the values of its endpoints in the following way.
If $f_0 \geq 0$ (which holds if and only if $\cc \leq \bb_1$), $\fk$ has no sign changes and a positive initial sign.
If $f_{n-1} \leq 0$ (which holds if and only if $\cc \geq \bb_n+ \rc (n-1) \Delta \bb_{n-1}$), $\fk$ has no sign changes and a negative initial sign.
If $f_0 < 0 < f_{n-1}$ (which holds if and only if $\bb_1 < \cc < \bb_n+ \rc (n-1) \Delta \bb_{n-1}$) $\fk$ has one sign change and a negative initial sign.
Result \ref{res:expressersconvex} follows from these observations upon applying \citet[result 3]{Pena2014}.

\subsection{Geometric benefits}
\label{subsec:expressersappgeometric}

For geometric benefits, we obtain the following result.

\begin{result}[Expresser-only traits with geometric benefits]
\label{res:expressersgeom}
Let $\fk$ be given by equation \eqref{eq:fkexpressers} with $\bb_k$ given by equation \eqref{eq:geombenefits} and let $\rc \geq 0$ and $n>2$ (the cases $\rc < 0$ or $n=2$ are trivial).
Moreover, let $\xi$, $\varsigma$ and $\tau$ be defined by equations \eqref{eq:xi} and \eqref{eq:varsigma}.
Then
\begin{enumerate}
\item If $\lambda \geq \xi$, $\Gg(z)$ is nondecreasing in $z$.
Furthermore
\begin{enumerate}
\item If $\cc/\bb \leq 1$, $z=1$ is the only stable point (A dominates B).
\item If $1 < \cc/\bb < \varsigma$, both $z=0$ and $z=1$ are stable and there is a unique internal unstable point $z^* \in (0,1)$ (bistability).
\item If $\cc/\bb \geq \varsigma$, $z=0$ is the only stable point (B dominates A).
\end{enumerate}
\item If $\lambda < \xi$, $\Gg(z)$ is unimodal in $z$, with mode given by $\hat{z}=\frac{1+\rc}{[1+\rc (n-1)](1-\lambda)}$.
Furthermore
\begin{enumerate}
\item If $\cc/\bb \leq 1$, $z=1$ is the only stable point (A dominates B).
\item If $1 < \cc/\bb \le \varsigma$, both $z=0$ and $z=1$ are stable and there is a unique internal unstable point $z^* \in (0,\hat{z})$ (bistability).
\item If $\varsigma < \cc/\bb < \tau$, there are two interior singular points $z_\mathrm{L}$ and $z_\mathrm{R}$ satisfying $z_\mathrm{L} < \hat{z} < z_\mathrm{R}$.
The points $z=0$ and $z_\mathrm{R}$ are stable, whereas $z_\mathrm{L}$ and $z=1$ are unstable (bistable coexistence).
\item If $\cc/\bb \geq \tau$, $z=0$ is the only stable point (B dominates A).
\end{enumerate}
\end{enumerate}
\end{result}

The arguments used for deriving this result are analogous to those used for deriving the results for nonexpresser-only traits with geometric benefits (result \ref{res:nonexpressersgeom} in app. \ref{sec:nonexpressersapp}).
Observing that $\xi <1$ holds and ignoring the trivial case $\lambda = 1$, there are three cases to consider: (i) $\lambda > 1$, (ii) $1 > \lambda \geq \xi$, and (iii) $\xi > \lambda$.

For $\lambda > 1$ synergies of scale are positive and hence result \ref{res:expressersconvex} applies with $\bb_1 = \bb$ and $\bb_n + \rc (n-1) \Delta \bb_{n-1} = \bb \varsigma$.
This yields result \ref{res:expressersgeom}.1 for the case $\lambda > 1$.

To obtain the results for the remaining two cases, we calculate the first and second forward differences of the benefit sequence \eqref{eq:geombenefits} and substitute them into
\begin{equation*}
 \Delta f_k = \Delta \bb_{k+1} + \rc \left\{ (k+1)\Delta^2 \bb_k + \Delta \bb_k \right\} , \ k = 0,1,\ldots,n-2 ,
\end{equation*}
to obtain
\[
  \Delta f_k = \bb \lambda^k \left[ \lambda(1+\rc) + \rc (\lambda-1) k \right] , \ k=0,1,\ldots,n-2 .
\]
For $\lambda < 1$, the sequence $\Dfk$ is decreasing in $k$ and hence can have at most one sign change.
Moreover, as $\Delta f_0 = \bb\lambda(1+\rc) > 0$ always holds true, the initial sign of $\Dfk$ is positive and whether or not the sequence $\Dfk$ has a sign change depends solely on how $\Delta f_{n-2}$ compares to zero.
Observe, too, that for $\lambda < 1$ we have $\varsigma > 1$ as $\lambda^n < \lambda$ holds.

Consider the case $\xi \leq \lambda < 1$. By the definition of $\xi$ (eq. \eqref{eq:xi}) this implies $\Delta f_{n-2} \geq 0$.
In this case $\Dfk$ has no sign changes and $\fk$ is nondecreasing.
The sign pattern of the inclusive gain sequence can then be determined by looking at how the signs of its endpoints depend on the cost-to-benefit ratio $\cc/\bb$.
If $\cc/\bb \leq 1$, then $f_0 \geq 0$, implying that $\fk$ has no sign changes and its initial sign is positive.
If $\cc/\bb \geq \varsigma$, then $f_n \leq 0$ and hence $\fk$ has no sign changes and its initial sign is negative.
If $1<\cc/\bb<\varsigma$, then $f_0 < 0 < f_n$, i.e., $\fk$ has one sign change and its initial sign is negative.
Result \ref{res:expressersgeom}.1 then follows from an application of \citet[result 3]{Pena2014}.

For $\lambda  < \xi$ we have  $\Delta f_{n-2} < 0$, implying that $\Dfk$ has one sign change from $+$ to $-$, i.e., $\fk$ is unimodal.
Hence, the gain function $\Gg(z)$ is also unimodal \citep[section 3.4.3]{Pena2014} with mode $\hat{z}$ determined by $\Gg'(\hat z) = 0$.
Using the assumption of geometric benefits, we can express $\Gg(z)$ is closed form as
\begin{equation*}
\Gg(z) = - \cc + \frac{\bb}{1-\lambda} + \bb \lambda \left\{\left[1+\rc (n-1)\right]z-\frac{1}{1-\lambda}\right\}(1-z+\lambda z)^{n-2},
\end{equation*}
with corresponding derivative
\begin{equation*}
\Gg'(z) = (n-1) \bb \lambda \left\{1+\rc -(1-\lambda)\left[1+\rc (n-1)\right]z\right\}(1-z+\lambda z)^{n-3}.
\end{equation*}
Solving $\Gg'(\hat z) = 0$ then yields $\hat z$ as given in equation \eqref{eq:hatzexpressers}.
The corresponding maximal value of the gain function is
\begin{equation*}
\label{eq:zghatexpressers}
\Gg(\hat{z}) = -\cc+\frac{\bb}{1-\lambda}\left[ 1+\lambda \rc \left( \frac{(n-2)\rc}{1+\rc (n-1)} \right)^{n-2} \right] .
\end{equation*}
Result \ref{res:expressersgeom}.2 then follows from applying \citet[result 5]{Pena2014}.
In particular, if $\cc/\bb \le 1$, we also have $\cc/\bb < \varsigma$, ensuring that $f_0 \ge 0$ and $f_{n-1} > 0$ hold (with unimodality then implying that the gain function is positive throughout).
If $1 < \cc/\bb \leq \varsigma$, we have $f_0 < 0$ and $f_{n-1} \ge 0$ (with unimodality then implying $\Gg(\hat z) > 0$).
If $\varsigma < \cc/\bb$, we have $f_0 < 0$ and $f_{n-1} < 0$.
Upon noticing that $\Gg(\hat{z})\leq 0$ is satisfied if and only if $\cc/\bb \geq \tau$ holds, this yields the final two cases in result \ref{res:expressersgeom}.2.

\newpage

\begin{table}
\begin{center}
  \begin{tabular}{ l | l }
    Symbol & Definition \\
    \hline
    A & first of two pure strategies (e.g., ``help'') \\
    $a_k$ & payoff to an A-player matched with $k$ A-players and $n-1-k$ B-players \\
    B & second of two pure strategies (e.g., ``do not help'') \\
    $B$ & payoff benefit parameter in games without synergies of scale \\
    $\Bb(z)$ & (fecundity) indirect effect \\
    $b$ & fitness benefit of carrying a social allele \\
    $b_k$ & payoff to a B-player matched with $k$ A-players and $n-1-k$ B-players \\
    $C$ & payoff cost parameter in games without synergies of scale \\
    $c$ & fitness cost of carrying a social allele \\
    $-\Cc(z)$ & (fecundity) direct effect \\
    $d_k$ & direct gain from switching to a focal matched with $k$ A-players and $n-1-k$ B-players \\
    $D$ & synergy parameter in games without synergies of scale \\
    $e_k$ & indirect gain from switching to a focal matched with $k$ A-players and $n-1-k$ B-players \\
    $f_k$ & inclusive gain from switching to a focal matched with $k$ A-players and $n-1-k$ B-players \\ 
    $\Gg(z)$ & gain function ($=-\Cc(z)+\kappa \Bb(z)$) \\
    $j$ & total number of A-players in a group \\
    $k$ & number of A co-players of a focal individual \\
    $N$ & number of adult individuals in a group \\
    $n$ & number of players (usually $= N$) \\
    $r$ & relatedness coefficient \\
    $\bb_j$ & benefit from the social good when $j$ individuals express helping (play A) \\
    $\bb$ & parameter of the geometric benefits \\
    $z$ & resident strategy (phenotype) \\
    $z_\bullet$ & strategy (phenotype) of a focal individual \\
    $z_{\ell(\bullet)}$ & strategy (phenotype) of the $\ell$-th co-player of the focal individual \\
    $z_\circ$ & average strategy (phenotype) of the neighbors of a focal individual \\ 
    $\Delta$ & first forward difference operator ($\Delta c_k= c_{k+1}-c_k$) \\
    $\gamma$ & payoff cost of expressing helping \\
    $\lambda$ & parameter of the geometric benefits \\
    $\rc$ & scaled relatedness coefficient \\
    $\pi$ & average payoff to a focal individual \\
  \end{tabular}
\end{center}
\caption{Symbols used in this article.}
\label{table:notation}
\end{table}

\begin{table}
\begin{center}
  \begin{tabular}{ l | l | l | l | l | l}
    Social trait               & $a_k$                 & $b_k$ & $d_k$          & $e_k$   & $f_k$        \\
    \hline
    whole-group         & $-\cc + \bb_{k+1}$    & $\bb_{k}$ & $-\cc\!+\!\Delta \bb_k$      & $(n\!-\!1) \Delta \bb_k$                          & $-\cc\!+\!(1 + \rc(n\!-\!1)) \Delta \bb_k$  \\
    \hline
    nonexpresser-only   & $-\cc$                & $\bb_k$  & $-\cc\!-\!\bb_k$              & $(n\!-\!1\!-\!k) \Delta \bb_k$                & $ -\cc\!-\!\bb_k\!+\!\rc (n\!-\!1\!-\!k) \Delta \bb_k$    \\
    \hline
    expresser-only      & $-\cc + \bb_{k+1}$    & $0$  & $-\cc\!+\!\bb_{k+1}$          & $k \Delta \bb_{k}$                            & $-\cc\!+\!\bb_{k+1}\!+\!\rc k \Delta \bb_{k}$ 
    \end{tabular}
\end{center}
\caption{Payoff structures and gains from switching for three kinds of social traits.
In each case expressers (A-players) incur a cost $\gamma > 0$ and recipients obtain a benefit $\bb_j \ge 0$ which depends on the number of expressers $j$ they experience.
The number of expressers experienced by a focal is $j = k$ if the focal is a nonexpresser, otherwise it is $j = k+1$.
Direct gains ($d_k$) and indirect gains ($e_k$) are calculated by substituting the expressions for $a_k$ and $b_k$ into equations~\eqref{eq:directgains} and~\eqref{eq:indirectgains}.
Inclusive gains from switching ($f_k$) are then obtained from equation~\eqref{eq:inclusivegains}.}
\label{table:threekinds}
\end{table}

\begin{table}
\begin{center}
  \begin{tabular}{ l | l | l }
    Social trait               & $f_k$                                     & $\Gg(z)$        \\
    \hline
    whole-group         & $-\cc\!+\!\bb\!+\!\rc(n\!-\!1) \bb$    &  $-\cc\!+\!\bb\!+\!\rc(n\!-\!1) \bb$  \\
    \hline
    nonexpresser-only   & $-\cc\!+\!\rc(n\!-\!1)\bb\!-\!(1\!+\!\rc)\bb k$                 & $-\cc\!+\!\rc(n\!-\!1)\bb\!-\!(1\!+\!\rc)\bb (n\!-\!1)z$ \\
    \hline
    expresser-only      & $-\cc\!+\! \bb\!+\!(1\!+\!\rc)\bb k$    & $-\cc\!+\! \bb\!+\!(1\!+\!\rc)\bb(n\!-\!1)z $
    \end{tabular}
\end{center}
\caption{Inclusive gains from switching ($f_k$) and gain functions ($\Gg(z)$) for the case of linear benefits (no synergies of scale).
The inclusive gains from switching are obtained by replacing $\bb_j = \bb j$ and $\Delta \bb_j = \bb$ into the corresponding expression from table \ref{table:threekinds}.
The gain function is then obtained from equation~\eqref{eq:Gzbinom}.
}
\label{table:linear}
\end{table}

\begin{table}
\begin{center}
  \begin{tabular}{| c | l | c | c | c |}
    \hline
    \multirow{4}{*}{whole-group}
    & \multicolumn{2}{|c|}{$\lambda < 1$} & \multicolumn{2}{|c|}{$\lambda > 1$} \\ \cline{2-5}
    & $\cc/\bb \leq \varepsilon$ & $z=1$ & $\cc/\bb \leq \varepsilon$ & $z=1$ \\ \cline{2-5}
    & $\varepsilon < \cc/\bb < \vartheta$ & $z^* \in (0,1)$ & $\varepsilon < \cc/\bb < \vartheta$ & $z=0$, $z=1$ \\ \cline{2-5}
    & $\cc/\bb \geq \vartheta$ & $z=0$ & $\cc/\bb \geq \vartheta$ & $z=0$ \\   
    \hline
    \hline
    \multirow{4}{*}{nonexpresser-only}
    & \multicolumn{2}{|c|}{$\lambda \leq \varrho$} & \multicolumn{2}{|c|}{$\lambda > \varrho$} \\ \cline{2-5}
    & $\cc/\bb < \zeta$ & $z^* \in (0,1)$ & $\cc/\bb < \zeta$ & $z^*\in(\hat{z},1)$ \\ \cline{2-5}
    & \multirow{2}{*}{$\cc/\bb \geq \zeta$} & \multirow{2}{*}{$z=0$} & $\zeta \leq \cc/\bb < \eta$ & $z=0$, $z_\textrm{R}\in (\hat{z},1)$ \\ \cline{4-5}
    & & & $\cc/\bb \geq \eta$ & $z=0$ \\
    \hline
    \hline
    \multirow{4}{*}{expresser-only}
    & \multicolumn{2}{|c|}{$\lambda < \xi$} & \multicolumn{2}{|c|}{$\lambda \geq \xi$} \\ \cline{2-5}
    & $\cc/\bb \leq 1$ & $z=1$ & $\cc/\bb \leq 1$ & $z=1$ \\ \cline{2-5}
    & $1 < \cc/\bb < \varsigma$ & $z=0$, $z=1$ & $1 < \cc/\bb < \varsigma$ & $z=0$, $z=1$ \\ \cline{2-5}
    & $\varsigma \leq \cc/\bb < \tau$ & $z=0$, $z_\textrm{R}\in (\hat{z},1)$ & \multirow{2}{*}{$\cc/\bb \geq \varsigma$} & \multirow{2}{*}{$z=0$} \\ \cline{2-3}
    & $\cc/\bb \geq \tau$ & $z=0$ &  & \\
    \hline
  \end{tabular}
\end{center}
\caption{Convergence stable strategies for
the three kinds of social traits with geometric benefits.
The results hold for whole-group traits if $\rc > -1/(n-1)$ and for nonexpresser-only and expresser-only traits if $\rc \geq 0$.
For whole-group traits,
$\varepsilon$ and $\vartheta$ are given by equation \eqref{eq:epsilonwhole}.
For nonexpresser-only traits, $\varrho$, $\zeta$, $\eta$, and $\hat{z}$ are functions of $\rc$, $n$, and $\lambda$ (see eq. \eqref{eq:varrho}, \eqref{eq:zeta}, and \eqref{eq:hatznonexpressers}).
For expresser-only traits, $\xi$, $\varsigma$, $\tau$, and $\hat{z}$ are functions of $\rc$, $n$, and $\lambda$ (see eq. \eqref{eq:xi}, \eqref{eq:varsigma} and \eqref{eq:hatzexpressers}).
}
\label{table:resultsfourkinds}
\end{table}

\bibliographystyle{elsarticle-harv}
\bibliography{gameswithrelatedness}

\newpage

\end{document}